\documentclass[10pt,conference,final]{IEEEtran}
\IEEEoverridecommandlockouts
\usepackage{cite}
\usepackage{amsmath,amssymb,amsfonts}
\usepackage{algorithmic}
\usepackage{graphicx}
\graphicspath{ {./results/} }
\usepackage{textcomp}
\usepackage{xcolor}
\def\BibTeX{{\rm B\kern-.05em{\sc i\kern-.025em b}\kern-.08em
    T\kern-.1667em\lower.7ex\hbox{E}\kern-.125emX}}

\usepackage[switch,columnwise]{lineno}

\usepackage{tikz}
\usetikzlibrary{arrows,decorations.pathmorphing,backgrounds,positioning,fit,petri}
\usetikzlibrary{automata}
\usetikzlibrary{graphs}
\usetikzlibrary{shadows}
\usetikzlibrary{shapes,arrows}

\usepackage[utf8]{inputenc}
\usepackage[T1]{fontenc}
\usepackage{grffile}
\usepackage{longtable}
\usepackage{wrapfig}
\usepackage{rotating}
\usepackage{textcomp}
\usepackage{amsthm}
\usepackage{capt-of}
\usepackage{hyperref}
\usepackage[margin=1in]{geometry}
\usepackage{multirow}
\usepackage{booktabs}
\usepackage{setspace}
\usepackage{bbding}

\newcommand*\BitAnd{\mathbin{\&}}
\newcommand*\BitOr{\mathbin{|}}
\newcommand{\BitNeg}{!}

\renewcommand{\arraystretch}{1.0}
\mathchardef\mhyphen="2D 

\newcommand{\cltloc}{CLTLoc}
\newcommand{\aez}{ae2sbvzot}
\newcommand{\oldTACK}{\textsc{ta2cltloc}}
\newcommand{\newTACK}{\textsc{ta2smt}}

\usepackage[textwidth=2.1cm,textsize=footnotesize]{todonotes}

\newcommand{\variable}{\ensuremath{n}}
\newcommand{\roland}{\textsc{Mitl$_{0,\infty}$BMC}}

\usepackage[final]{changes}
\definechangesauthor[name={MB},color=blue]{MB}
\newcommand{\addMB}[1]{\added[id=MB]{#1}}

\newcommand{\repMB}[2]{\replaced[id=MB]{#2}{#1}}

\newcommand*{\intNum}{\ensuremath{\mathbb{Z}}}
\newcommand*{\naturalNum}{\ensuremath{\mathbb{N}}}
\newcommand*{\realNum}{\ensuremath{\mathbb{R}}}
\newcommand*{\partSet}{\wp}

\newcommand{\transitions}[1]{\ensuremath{T_{#1}}}
\newcommand{\netid}{\ensuremath{\mathcal{N}}}

\newcommand{\clock}{\ensuremath{x}}
\newcommand{\clocks}{\ensuremath{X}}

\newcommand{\taid}{\ensuremath{\mathcal{A}}}

\newcommand{\constant}{\ensuremath{d}}

\newcommand{\variables}{\ensuremath{\mathit{Int}}}

\newcommand{\clockconstraint}{\ensuremath{\gamma}}
\newcommand{\variableconstraint}{\ensuremath{\xi}}
\newcommand{\varassignement}{\ensuremath{\mu}}
\newcommand{\nullevent}{\ensuremath{\tau}}
\newcommand{\actions}{\ensuremath{Act_\nullevent}}
\newcommand{\action}{\ensuremath{\alpha}}
\newcommand{\autstate}{\ensuremath{q}}

\newcommand{\transtavar}{\ensuremath{q \xrightarrow{\clockconstraint, \variableconstraint, \action, \resettedclocks, \varassignement} q^\prime}}
\newcommand{\conftranstavar}[1]{\ensuremath{l[ #1 ] \xrightarrow{\clockconstraint, \variableconstraint, \action, \resettedclocks, \varassignement} l^\prime[ #1 ]}}

\newcommand{\tanumber}{\ensuremath{N}}

\newcommand{\loc}{\mathtt{l}}
\newcommand{\tracesymbol}{\ensuremath{\eta}}
\newcommand{\signalsymbol}{\ensuremath{M_\tracesymbol}}

\newcommand{\proposition}{\ensuremath{p}}

\newcommand{\autindex}{\ensuremath{i}}

\newcommand{\numberOfTA}{\ensuremath{K}}

\newcommand{\ie}{\mathtt{ie}}
\newcommand{\ei}{\mathtt{ei}}

\newcommand{\clockvaluationfunction}{\ensuremath{v}}

\newcommand{\variablevaluationfunction}{\ensuremath{v_{\mathrm{var}}}}
\newcommand{\variablevaluationfunctionpar}[1]{\ensuremath{v_{\mathrm{var},#1}}}
\newcommand{\clockvaluationfunctionpar}[1]{\ensuremath{v}_{#1}}

\newcommand{\resettedclocks}{\ensuremath{\zeta}}

\newcommand{\eventtime}{\ensuremath{\Upsilon}}

\newtheorem{definition}{Definition}

\begin{document}

\title{Improved Bounded Model Checking\\of Timed Automata
\thanks{2021  IEEE.  Personal  use  of  this  material  is  permitted.  Permissionfrom IEEE must be obtained for all other uses, in any current or futuremedia,  including  reprinting/republishing  this  material  for  advertisingor promotional purposes, creating new collective works, for resale orredistribution to servers or lists, or reuse of any copyrighted componentof this work in other works.}
}

\author{
\IEEEauthorblockN{Robert L. Smith\IEEEauthorrefmark{1}, Marcello M. Bersani\IEEEauthorrefmark{1}, Matteo Rossi\IEEEauthorrefmark{2}, Pierluigi {San Pietro}\IEEEauthorrefmark{1}}
\IEEEauthorblockA{\IEEEauthorrefmark{1}Dipartimento di Elettronica, Informazione e Bioingegneria, Politecnico di Milano. Milano, Italy.}
\IEEEauthorblockA{\IEEEauthorrefmark{2}Dipartimento di Meccanica, Politecnico di Milano. Milano, Italy.\\
Email: \texttt{robert.smith@mail.polimi.it},
\\\{\texttt{marcellomaria.bersani, matteo.rossi, pierluigi.sanpietro}\}@\texttt{polimi.it}}
}

\maketitle

\begin{abstract}
Timed Automata (TA) are a very popular modeling formalism for systems with time-sensitive properties. 
A common task is to verify if a network of TA satisfies a given property,  
usually expressed in Linear Temporal Logic (LTL), or in a subset of Timed Computation Tree Logic (TCTL).
In this paper, we build upon the TACK bounded model checker for TA, which supports 
a signal-based semantics of TA and 
the richer Metric Interval Temporal Logic (MITL).
TACK encodes both the TA network and property into a variant of LTL, Constraint LTL over clocks (\cltloc).
The produced \cltloc\ formula can then be solved by tools such as Zot, which transforms \cltloc\ properties into 
the input logics of Satisfiability Modulo Theories (SMT) solvers.
%
We present a novel method that preserves TACK's encoding of MITL properties
while encoding the TA network directly into the SMT solver language, making use of
both the BitVector logic and the logic of real arithmetics. 
We also introduce several optimizations that allow us to significantly outperform the \cltloc\ encoding in many practical scenarios.  
\end{abstract}


\section{Introduction}
\label{sec:Intro}

Timed Automata~\cite{alur94} (TA) are a popular tool for modeling time-sensitive systems.
By combining the transition semantics of finite state automata with real-valued clocks, they are of great theoretical and practical interest for representing time-bound processes and applications.
They have found common use in the domain of model checking, where system representations are evaluated against a given property of interest.
Various tools and languages exist for a variety of applications and use cases.
These include the current de facto standard Uppaal~\cite{larsen97}, as well as NuSMV~\cite{cimatti02}.

Model Checking refers to a verification technique for solving properties of state transition systems.
A wide variety of industrial applications, including circuit design, control systems, and program verification lend themselves to
this representation.
In the model checking process, the system is exhaustively searched to see if the given property is valid.
%
TACK is a bounded model checker for networks of TA \cite{tack20}. 
Properties to be verified are specified in Metric Interval Temporal Logic (MITL) \cite{AFH96}, and are converted along with the TA network into \cltloc \cite{bersani2016tool}, a variant of Linear Temporal Logic (LTL) supporting real-valued clocks.

This paper presents a novel encoding of the TA network which does not use \cltloc\ as an intermediate step, instead directly transforming the network semantics into a hybrid BitVector representation.
This approach has the advantage of being tailor-made for TA networks, while the previous approach relied on the general-purpose \cltloc\ converter \aez{} \cite{PRB20}.
However rather than just re-create the existing encoding in a new language, we have corrected several
deficiencies in the original TACK encoding, and have introduced new features to make TACK more useful for users.
We have also exploited opportunities to more efficiently encode TA constructs, noticeably eliminating the need for BitVectors to track the active state of the TA, instead relying on the active transition to carry this information.

In this paper, we first present the current state-of-the-art for bounded model checking, followed by an in-depth description of both the required preliminary knowledge and the specific implementation of the TACK bounded model checker (Section~\ref{sec:Prelim}).
We then introduce our novel encoding of TA networks into a form suitable for an SMT-based bounded model checker (Section \ref{sec:ImprovEnc}, and we present experimental results comparing the new encoding with existing ones (Section \ref{sec:Exp}).
Finally, we conclude with a discussion of the result and some future works (Section \ref{sec:Concl}).

\section{Preliminaries}
\label{sec:Prelim}

\subsection{State of the Art}
\label{sota}

For many years, model checking was performed using Binary Decision Diagrams (BDDs)~\cite{bryant86}, which offer many time- and space-complexity advantages over explicit state enumeration~\cite{burch92}.
However to efficiently handle larger state spaces, bounded model checking techniques techniques have been developed.
Bounded model checking encodes the verification problem of the state transition system into a propositional satisfiability (SAT) or Satisfiability Modulo Theories (SMT) problem, and then tasks the SAT/SMT solver with finding a valid assignment of states to time positions starting from a given initial state such that the desired property is violated (counterexample); if no such assignment is found, the property holds for the system.
Because such solvers require finite state spaces, the number of time positions considered is limited by a bound $k$, hence the name bounded model checking.
%
Bounded model checking analyzes traces of infinite length that can be represented in finite space.
This is accomplished by limiting the search to so-called ``lasso-shaped'' traces.
These traces begin with an initial finite sequence of states before entering an infinite loop of states.
Thus only a finite number of states need to be explicitly represented by the bounded model checker, which can search for lassos of length up to the given bound.

%
Uppaal~\cite{larsen97} is a de facto standard for model checking systems modeled through TA.
Uppaal allows users to express properties to be checked using Timed Computation Tree Logic (TCTL), an extension of Computation Tree Logic (CTL) with real-time properties \cite{baier2008principles}.
However, Uppaal and similar implementations restrict themselves to only a subset of TCTL, which focuses mostly on reachability and invariant properties.

In addition to the work done with branching-time logics, there has been interest in combining TA with the expressive power of Metric Temporal Logic (MTL), an extension of LTL with interval constraints on the `until' operator \cite{FMMR12}.
While powerful, MTL is undecidable in general for infinite traces~\cite{bouyer09}.
MITL \cite{AFH96} is a decidable restriction of MTL which can capture more complex properties than those supported by the subset of TCTL allowed by TA model checkers.
In recent years bounded model checkers supporting MITL as property specification language have been developed, in particular \roland{} \cite{kindermann13}, MightyL \cite{brihaye2017m} and TACK (see \cite{tack20} for a detailed comparison of the tools).
In this work we improve the encoding of the TA verification problem into an SMT problem used in TACK.

\subsection{Timed Automata}
\label{timed-automata}

\newcommand{\channel}{\mathit{channel}}
\newcommand{\sync}{\mathit{sync}}
\newcommand{\Int}{\mathit{Int}}
\newcommand{\Exp}{\mathit{exp}}
\newcommand{\Inv}{\mathit{Inv}}

Timed Automata (TA) are a popular formalism for modeling interactions that require precise timing mechanisms~\cite{alur94}. 
In this paper, we consider an extension of TA that includes integer variables with finite ranges and mechanisms to synchronize the taking of transitions. 

Let \(AP\) be a set of atomic propositions, and let \(Act\) be a set of synchronization events of the form $Act \subset \{\channel \times \sync\}$, where $\channel$ is a finite set of symbols and $\sync \in \{!,?,\#,@\}$.
In addition we define a null event \(\tau\).
\(Act_{\tau}\) is the set \(Act \cup \{\tau\}\).
Let \(X\) be a finite set of clocks, and \(\Int\) a finite set of integer-valued variables.
\(\Gamma(X)\) is the set of clock constraints, where a clock constraint \(\gamma\) is a relation
\(x \sim c \BitOr \gamma \land \gamma\), where \(x \in X\), \(\sim \in \{<,>,\leq,\geq\}\), and \(c \in \mathbb{N}\).
\(Assign(X)\) is the set of clock assignments, where each assignment has the form \(x := 0\), where \(x \in X\). \(Assign(\Int)\) is a set of variable assignments of the form \(y := \Exp\), where \(\Exp := \Exp + \Exp\BitOr \Exp - \Exp\BitOr n\BitOr c\), \(n \in \Int\) and \(c \in \mathbb{Z}\). \(\Gamma(\Int)\) is the set of integer variable constraints, where a variable constraint \(\gamma\) is defined as
\(\gamma := n \sim c\BitOr n \sim n'\BitOr \neg \gamma\BitOr \gamma \land \gamma\), where \(n\) and \(n'\) are integer variables, \(c \in \mathbb{Z}\), and \(\sim \in \{<,=\}\).

A TA with variables is defined as the tuple \(\mathcal{A} = \big \langle AP,X, Act_{\tau}, \Int, Q, q^0, v_\mathrm{var}^0, \Inv, L, T \big \rangle\), where
\(Q\) is a finite set of locations, \(q^0 \in Q\) is the initial location, \(v_\mathrm{var}^{0} : \Int \rightarrow \mathbb{Z}\) is a function providing initial values for each of the variables, and \(\Inv : Q \rightarrow \Gamma(X)\) is a function assigning each location to a (possibly empty) set of clock constraints, which are the invariants of the location.
The labeling function \(L: Q \rightarrow \partSet(AP)\) assigns each location to a subset of the atomic propositions. Each transition \(t \in T\) has the form \(t = \big \langle Q \times Q \times Act_{\tau} \times \Gamma(X) \times \Gamma(\Int) \times \partSet(Assign(X)) \times \partSet(Assign(\Int)) \big \rangle \), consisting of a source and destination location, an action, a set of clock and variable guards, a set of clocks to be reset when the transition fires, and a set of variables to assign values to. To refer to the components of a transition we will use \(t_-\) and \(t_+\) to refer to the source and destination locations respectively, as well as
\(t_\epsilon, t_{\gamma_c}, t_{\gamma_v}, t_{a_c}, t_{a_v}\) to refer to the event, clock constraints, variable constraints, clock assignments, and variable assignments respectively.
{A transition is written as \transtavar{}, where 
\clockconstraint\ is a constraint of $\Gamma(X)$,
\variableconstraint\ is a constraint of $\Gamma(\Int)$, $\alpha$ is an element of $Act_\tau$,
$\zeta$ is a subset of $X$ and \varassignement\ is {a set of assignments} from
$\partSet(Assign(\Int))$.
Let $U(\mu)$ be the set of variables that are updated by $\mu$---that is, that appear as the left-hand side in an assignment of $\mu$---and let $U(t)$ indicate the set $U(\mu)$ given a transition $t$.
}

\begin{figure}[h]
  \centering
  \includegraphics[width=0.35\textwidth]{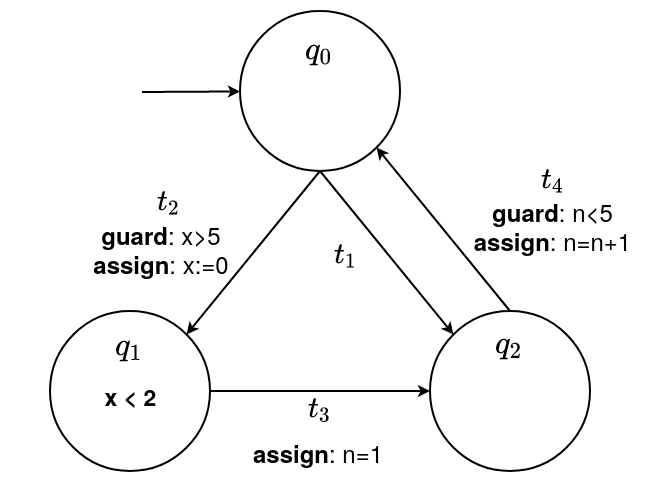}
  \caption{A Timed Automaton with clock $x$ and variable $n$.
  }
  \label{fig:example-big}
\end{figure}

We outline the semantics of networks of TA, and we illustrate its key features through the simple example shown in Figure \ref{fig:example-big}; we then show the formal definition.
%
Transition guards are conditions over either clocks or variables that prevent the associated transition from being taken when they are not satisfied.
As an example, transition $t_{2}$ can only be taken when the value of clock $x$ is greater than $5$.
Assignments on the other hand modify the value of a clock or variable \emph{after} the transition has been taken.
For example, it is 
valid for transition $t_{2}$ to be taken when $x=6$, even though the assignment $x:=0$ resets the value of $x$ to $0$. 
{A transition is said to be \emph{enabled} if the values of clocks and variables satisfy the guard, and \emph{active} at the time when it is fired.}
The value is updated in the same instant as the transition, however the guards only consider the pre-transition values of the clocks when determining if the transition is valid.
Variables can be assigned to any value, while clocks can only be reset to $0$.
When a TA is in a certain location, the corresponding invariant (if any) is required to be true.
The invariant attached to $q_{2}$ requires the TA to leave location $q_{2}$ before clock $x$ reaches a value of $2$.

\begin{figure}
\scalebox{0.7}{
	\begin{tikzpicture}
	\draw node at (0,0) (a) {\small $c_0$};
	\draw node at (3,0) (b) {\small $c'_0$};
	\draw node at (3,0.8) (c) {\small $c_1$};
	\draw node at (7,0.8) (d) {\small $c'_1$};
	\draw node at (7,0) (e) {\small $c_2$};
	\draw node at (9,0) (f) {\small $c'_2$};
	\draw node at (9,0.8) (g) {\small $c_3$};
	\draw[->] (a) -- (b) node[midway,above]{};
	\draw[->] (b) -- (c) node[midway,left]{\footnotesize $t_1$};;
	\draw[->] (c) -- (d) node[midway,above]{};
	\draw[->] (d) -- (e) node[midway,left]{\footnotesize $t_4$};;
	\draw[->] (e) -- (f) node[midway,above]{};
	\draw[->] (f) -- (g) node[midway,left]{\footnotesize $t_2$};
	\draw node at (0.5,0)  {\tiny $|$};
	\draw node at (0.5,0)  {\tiny $|$};
	\draw node at (1.7,0)  {\tiny $|$};
	\draw node at (2,0)  {\tiny $|$};
	\draw node at (5,0.776)  {\tiny $|$};
	\draw node at (6,0.776)  {\tiny $|$};
	\draw node at (8.5,0)  {\tiny $|$};
	%
	%
	
	\pgfmathsetmacro{\ilocation}{-0.75}
	\draw[dashed] (-1,-0.5) -- (9.3,-0.5);
	\pgfmathsetmacro{\llocation}{-1}
	\pgfmathsetmacro{\dlocation}{-1.5}
	\pgfmathsetmacro{\xlocation}{-2.0}
	\pgfmathsetmacro{\tlocation}{-2.0}
	\pgfmathsetmacro{\clocation}{-3.1}
	\pgfmathsetmacro{\elocation}{-3.6}
	\draw node at (-1,\llocation) {\small $p=$};
	\draw node at (-1,\dlocation) {\small $\variable=$};
	\draw node at (-1,\tlocation) {\small $t=$};
	%
	\draw node at (0,\llocation) {\small $q_0$};
	\draw node at (0.5,\llocation) {\small $q_0$};
	\draw node at (1.7,\llocation) {\small $q_0$};
	\draw node at (2,\llocation) {\small $q_0$};
	\draw node at (3,\llocation) {\small $q_2$};
	\draw node at (5,\llocation) {\small $q_2$};
	\draw node at (6,\llocation) {\small $q_2$};
	\draw node at (7,\llocation) {\small $q_0$};
	\draw node at (8.5,\llocation) {\small $q_0$};
	\draw node at (9,\llocation) {\small $q_1$};
	\draw node at (0,\dlocation) {\small $0$};
	\draw node at (0.5,\dlocation) {\small $0$};
	\draw node at (1.7,\dlocation) {\small $0$};
	\draw node at (2,\dlocation) {\small $0$};
	\draw node at (3,\dlocation) {\small $0$};
	\draw node at (5,\dlocation) {\small $0$};
	\draw node at (6,\dlocation) {\small $0$};
	\draw node at (7,\dlocation) {\small $1$};
	\draw node at (8.5,\dlocation) {\small $1$};
	\draw node at (9,\dlocation) {\small $1$};

	\draw node at (0,\tlocation) {\small $\sharp$};
	\draw node at (0.5,\tlocation) {\small $\sharp$};
	\draw node at (1.7,\tlocation) {\small $\sharp$};
	\draw node at (2,\tlocation) {\small $t_1$};
	\draw node at (3,\tlocation) {\small $\sharp$};
	\draw node at (5,\tlocation) {\small $\sharp$};
	\draw node at (6,\tlocation) {\small $t_4$};
	\draw node at (7,\tlocation) {\small $\sharp$};
	\draw node at (8.5,\tlocation) {\small $t_2$};
	\draw node at (9,\tlocation) {\small $\sharp$};

	%
	%
	\draw[dashed] (-1,-2.5) -- (9.3,-2.5);
	\draw node at (-1,\elocation) {\small $\mathtt{edge}=$};	
	\draw node at (0,\clocation)  {$[$};
	\draw node at (2.98,\clocation)  {$]$};
	\draw[-] (3,\clocation) -- (0,\clocation) node[midway,above]{\footnotesize $c_0$};
	\draw node at (3.02,\clocation)  {$($};
	\draw node at (6.98,\clocation)  {$)$};
	\draw[-] (7,\clocation) -- (3,\clocation) node[midway,above]{\footnotesize $c_1$};
	\draw node at (7.02,\clocation)  {$[$};
	\draw node at (8.98,\clocation)  {$]$};	
	\draw[-] (9,\clocation) -- (7,\clocation) node[midway,above]{\footnotesize $c_2$};			
	\draw node at (0,\elocation) {$\cdot$};
	\draw node at (0.5,\elocation) {$\cdot$};
	\draw node at (1.7,\elocation) {$\cdot$};
	\draw node at (2,\elocation) {$\cdot$};
	\draw node at (3,\elocation) {$]($};
	\draw node at (5,\elocation) {$\cdot$};
	\draw node at (6,\elocation) {$\cdot$};
	\draw node at (7,\elocation) { $)[$};
	\draw node at (8.5,\elocation) {$\cdot$};
	\draw node at (9,\elocation) { $]($};		
	\end{tikzpicture}
	}
	\caption{
Illustration of the semantics of the TA of Fig.~\ref{fig:example-big}.}
	\label{fig:differentsemantics}
\end{figure}
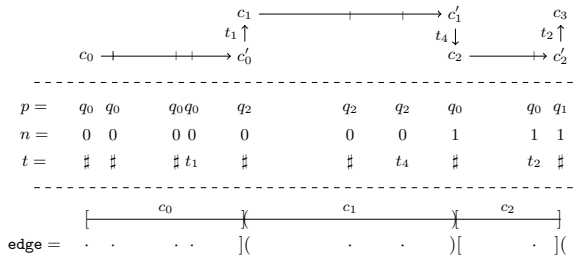 


\begin{definition}\label{def:ta-config}
Given a TA $\mathcal{A}$, a \emph{configuration} of $\mathcal{A}$ is a tuple $( q, v_{var}, v )$ where $q$ is the current location of $\mathcal{A}$ and $v_{var}$ (resp., $v$) is a variable (resp., clock) valuation $\Int \rightarrow \mathbb{Z}$ (resp., $X \rightarrow \mathbb{R}_{\geq 0}$).
\end{definition}
We adopt a semantics for TA based on so-called \emph{signals}, where each instant of the time domain $\mathbb{R}_{\geq 0}$ comprising all nonnegative real numbers is associated with a configuration.
The configuration of a TA changes when a transition is taken, but it does not change between transitions.
Hence, we can split the time domain $\mathbb{R}_{\geq 0}$ into intervals during which the configuration of the TA remains the same.
Figure \ref{fig:differentsemantics} shows a fragment of an execution of the TA of Figure \ref{fig:example-big}.
The location is initially $q_0$ (in configuration $c_0$), then it changes to $q_2$ (and configuration $c_1$) when transition $t_1$ is taken.
As Figure \ref{fig:differentsemantics} shows, in the instant in which a transition is taken the configuration can be the old or the new one, depending on whether the \emph{edge} of the transition is right-closed ($]($) or left-closed ($)[$).
For example, in Figure \ref{fig:differentsemantics} the switch from configuration $c_0$ to $c_1$ occurs in a right-closed manner, whereas the one between $c_1$ and $c_2$ in a left-closed one.

A network of TA is a finite set of TA \(\mathcal{N} = [\mathcal{A}_1, \mathcal{A}_2, \ldots \mathcal{A}_\tanumber]\).
TA in the same network can refer to common clocks, variables, and synchronization channels to coordinate their actions.
To simplify the notation we will use the symbols \(T\), \(X\), \(\Int\), and \(Act/Act_{\tau}\) to refer to the union of the respective sets of each individual TA in the network.
When necessary to refer to the properties of one timed automaton in particular, we will append a numerical subscript to the set in question, for example \(X_i\) to refer to the clocks used by the specific timed automaton \(\mathcal{A}_{i} \in \mathcal{N}\).

{
Before providing the formal definition of the transition relation for networks of TA, 
the notion of \emph{weak} satisfaction relation $\models_w$ over clock valuations and clock constraints is introduced, where $\sim\in\{<,>,\leq,\geq\}$.
\begin{center}
\begin{tabular}{lll}
$v \models_w x \sim d$ & iff \ \ \ $v(x) \sim d \text{ or } v(x) = d$ &  \\
$v \not \models_w x = d$ & for any $x \in X, d \in \mathbb{N}.$ 
\end{tabular}
\end{center}
Naturally, $\models_w$ can be extended to conjunctions of formulae $x \sim \constant$.
For instance, the formula {$x<1 \land \neg(y<1) \land \neg(y=1)$} is both satisfied and weakly satisfied by the clock evaluation such that $v(x)=0.8$ and $v(y)=1.2$, but it is only weakly satisfied if $v(x)=1$ and $v(y)=1$.

\newcommand{\nillevent}{\tau}
\begin{definition}
Let \netid\  be a network of \tanumber\ TA $\taid_1, \ldots, \taid_\tanumber$.
A \emph{configuration} of \netid\ is a tuple $( \loc , \variablevaluationfunction, \clockvaluationfunction )$ where $\loc$ is a vector $[\autstate^1, \ldots, \autstate^\tanumber]$ 
such that
$\autstate^1, \ldots, \autstate^\tanumber$ are locations of 
$\taid_1, \ldots, \taid_\tanumber$, and \variablevaluationfunction\ (resp., \clockvaluationfunction) is a variable (resp., clock) valuation for the set
\variables\ (resp.,  \clocks )
 including all integer variables  (resp., clocks) appearing in $\taid_1, \ldots, \taid_\tanumber$.  
\end{definition} 
 
When a network of TA is considered,  it is possible that some automata in the network take a transition while the remaining others do not fire a transition and keep their state unchanged.
Firing a transition labeled with the null event \nullevent\ (i.e., a transition that does not synchronize, as explained later) is different from not taking a transition at all.
Symbol $\_$ indicates that an automaton $\taid_\autindex$ does not perform any transition in \transitions{\autindex}.
The notation $\loc[\autindex]$ indicates the location of automaton $\taid_\autindex$---i.e., if $\loc[\autindex] = j$, then automaton $\taid_\autindex$ is in location $\autstate^\autindex_j$, assuming that the locations of each automaton are numbered, with $0$ indicating the initial one. 
The two kinds of configuration changes that may occur when an automaton in the network performs a transition from a location $q$ to $q^\prime$ are indicated in Def.~\ref{def:confatrans} with symbols $\ei$ ({excluded-included, or left-closed}) and $\ie$ (included-excluded, or right-closed).

	\begin{definition}
		\label{def:confatrans}
		Let \netid\  be a network of \tanumber\ TA.
		Let $(\loc, \variablevaluationfunction, \clockvaluationfunction)$, $(\loc', \variablevaluationfunction', \clockvaluationfunction')$ be two configurations, let $\delta \in \realNum_{>0}$ and $\Lambda$ be a tuple of \tanumber\ symbols such that $\Lambda[k] \in \{\actions \times \{\ei, \ie\}\} \cup \{ \_  \}$
		for every $1 \le \autindex \leq \tanumber$. 
		Then, a configuration change is either a transition 
		$(\loc, \variablevaluationfunction, \clockvaluationfunction) \xrightarrow{\Lambda} (\loc', \variablevaluationfunction', \clockvaluationfunction')$ 
		or  a transition 
		$(\loc, \variablevaluationfunction, \clockvaluationfunction) \xrightarrow{\delta} (\loc', \variablevaluationfunction', \clockvaluationfunction')$ defined as follows.
		\begin{enumerate}
			\item 				\label{condone} $(\loc, \variablevaluationfunction, \clockvaluationfunction) \xrightarrow{\Lambda} (\loc', \variablevaluationfunction', \clockvaluationfunction')$  occurs if 
			\begin{enumerate}
				\item 
				\label{firing}
				for each $\Lambda[\autindex]=(\action, b)$
				there is a transition 
				$\conftranstavar{\autindex}$	 in $\mathcal{A}_{\autindex}$ such that:
				\begin{enumerate}
					\item
					\label{guards}
					$\clockvaluationfunction \models \clockconstraint$ and $\variablevaluationfunction \models \variableconstraint$, 

					\item
					\label{resets}
					$\clockvaluationfunction'(\clock)=0$ holds for all $\clock \in \resettedclocks$, 

					\item
					\label{assignments}
					$(\variablevaluationfunction',\variablevaluationfunction) \models \varassignement$, 

					\item
					\label{invariant-leftopenrightclosed}
					when $b = \ei$ then:
					\begin{itemize}
						\item $\clockvaluationfunction \models_w \Inv(\loc[\autindex])$ and
						\item $\clockvaluationfunction' \models \Inv(\loc'[\autindex])$
					\end{itemize}
					\item\label{invariant-leftclosedrightopen}
					when $b = \ie$ then:
					\begin{itemize}
						\item $\clockvaluationfunction \models \Inv(\loc[\autindex])$ and 
						\item $\clockvaluationfunction' \models_w \Inv(\loc'[\autindex])$ 
					\end{itemize}
				\end{enumerate}
				\item
				\label{keepstate} 
				for each $\Lambda[\autindex] = \_$ it holds that:
				\label{epsilon-transition}
				\begin{enumerate}
					\item
					\label{epsilon-keepstate} 
					$\loc'[\autindex]=\loc[\autindex]$;
					\item $\clockvaluationfunction \models \Inv(\loc[\autindex])$ and $\clockvaluationfunction' \models \Inv(\loc'[\autindex])$.
				\end{enumerate}
				\item
				\label{keepclocksandvariables}
				for each clock $\clock \in \clocks$ (resp., integer variable $\variable \in \variables$), if \clock\ (resp., \variable) does not appear in any \resettedclocks\
				(resp., it is not assigned by any $A$) of one of the transitions taken by $\mathcal{A}_{1}, \ldots, \mathcal{A}_{\tanumber}$, then $\clockvaluationfunction'(x) = \clockvaluationfunction(x)$ (resp., $\variablevaluationfunction'(n)=\variablevaluationfunction(n)$);
			\end{enumerate}
			\item \label{time-transition} $(\loc, \variablevaluationfunction, \clockvaluationfunction) \xrightarrow{\delta} (\loc', \variablevaluationfunction', \clockvaluationfunction')$ 
occurs if 
			$\loc'=\loc$,
			$\variablevaluationfunction'= \variablevaluationfunction$,
			$\clockvaluationfunction'= \clockvaluationfunction+\delta$
			and for all $1 \leq \autindex \leq K$,
			$\clockvaluationfunction' \models_w \Inv(\loc[\autindex])$.
		\end{enumerate}
	\end{definition}

A configuration change $(\loc, \variablevaluationfunction, \clockvaluationfunction ) \xrightarrow{\Lambda} (\loc', \variablevaluationfunction', \clockvaluationfunction')$, for some $\Lambda \in  \{\{\actions \times \{\ei, \ie\}\} \cup \{ \_  \}\}^\numberOfTA$,
satisfying (1) is called a \emph{discrete transition}.
If it satisfies (2) then it is called a \emph{time transition}.
For convenience of notation, symbols $(\action,\ei)$ and $(\action,\ie)$, for some $\action \in \actions$, are hereinafter denoted, respectively, with $\action^{)[}$ and $\action^{](}$.
	The edge of a transition realized with an action $\action$ is determined by the conditions in~\ref{invariant-leftopenrightclosed} and~\ref{invariant-leftclosedrightopen} and depend on the invariants of the locations involved in the transition, the clock values and the resets applied in the configuration change. 
	Cases~\ref{condone}) and~\ref{time-transition})  are discussed in detail in \cite{tack20}.
	
The notions of trace and signal are now introduced.

\begin{definition}
\label{def:trace}
Let \netid\  be a network of \tanumber\ TA. 
A \emph{trace} of \netid{} is an infinite sequence $\eta$ of the form \[(\loc_0, \variablevaluationfunctionpar{0}, \clockvaluationfunctionpar{0}), {e_0}, (\loc_1, \variablevaluationfunctionpar{1}, \clockvaluationfunctionpar{1}), {e_1}, \ldots\] such that:
\begin{enumerate}
\item for all $h \in \naturalNum$, $e_h = \Lambda_h$ or $e_h = \delta_h$;
\item
{
for all $h \in \naturalNum$ it holds that $(\loc_h, \variablevaluationfunctionpar{h}, \clockvaluationfunctionpar{h})  \xrightarrow{e_h} (\loc_{h+1},\variablevaluationfunctionpar{h+1}, \clockvaluationfunctionpar{h+1})$;}

\item \label{firstTransitionIsDelay}
$e_0 = \delta_0$, for some $\delta_0 \in \realNum_{>0}$;

\item
\label{invariantholdsinstate}
for all $1 \leq \autindex \leq \tanumber$, it holds that $\loc_0[\autindex]=0$, $\clockvaluationfunction_0 \models \Inv(\loc_0[\autindex])$, for all $\clock \in \clocks$ it holds that $\clockvaluationfunctionpar{0}(\clock) = 0$, and for all $\variable \in \variables$ it holds that $\variablevaluationfunctionpar{0}(n) = \variablevaluationfunction^0(\variable)$.

\item
\label{noconsecLambda}
discrete transitions must be followed by time transitions; that is, if {$e_h$} is a discrete transition ({$e_h = \Lambda_{h}$}), then {$e_{h+1}$} is a time transition ({$e_h = \delta_{h+1}$}).
 
\end{enumerate}
\end{definition}


Since by condition \ref{noconsecLambda} there cannot be two consecutive discrete transitions, and since any finite sequence of consecutive delays $\delta_h \dots \delta_{h+k}$, with $k\geq 0$, is equivalent to a single delay $\sum_{j=h}^{h+k} \delta_j$, a trace can always be rewritten into
a new one such that discrete and time transitions strictly alternate.
Moreover, by the previous property, every {time transition $\delta_h$ can be replaced with}  a finite
sequence of $m$ pairs of time and discrete transitions $\delta_{h,0}\Lambda_{h,0} \delta_{h,1}\Lambda_{h,2} \dots \delta_{h,m-1}$, strictly alternating, such that $\Lambda_{h,j}[\autindex]=\_$ holds for all $0\leq j\leq m-1$, $1 \leq \autindex \leq \tanumber$, and $\delta_{h} = \sum_{j=0}^{m-1} \delta_{h,j}$.

With a slight abuse of notation, a trace is represented in the following way, where the numbering of configurations increases only after discrete transitions: 


\begin{align*}
(\loc_0, \variablevaluationfunctionpar{0}, \clockvaluationfunctionpar{0}) 
\xrightarrow{\delta_0} \ &
(\loc'_0, \variablevaluationfunctionpar{0}', \clockvaluationfunctionpar{0}') 
\xrightarrow{\Lambda_0} \\
& (\loc_1, \variablevaluationfunctionpar{1}, \clockvaluationfunctionpar{1}) 
\xrightarrow{\delta_1}  \ldots  \nonumber
\end{align*}

Traces encode executions of TA by means of denumerable sequences of time and discrete transitions.
However, the evolution of a network of TA is continuous, hence it is more naturally represented by means of \emph{signals}.
Intuitively, given a trace \tracesymbol, the projection over the real line of the values of its integer variables and atomic propositions associated with locations determines a signal \signalsymbol.
To be able to consistently associate signals with traces of a TA, however, we impose the following restriction on traces.

\begin{definition}
\label{rem:tracerestriction}
Let $\netid$ be a network of TA. A trace $\eta$ of $\netid$ is \emph{edge-consistent} if, for any configuration change $(\loc'_h, \variablevaluationfunctionpar{h}', \clockvaluationfunctionpar{h}') 
\xrightarrow{\Lambda_h}
(\loc_{h+1}, \variablevaluationfunctionpar{h+1}, \clockvaluationfunctionpar{h+1})$
there are two transitions $ \loc'_h[ \autindex ] \xrightarrow{\clockconstraint, \variableconstraint, \action, \resettedclocks, \varassignement} \loc_{h+1}^\prime[ \autindex ]$ and $\loc'_h[ \bar{\autindex} ] \xrightarrow{\bar{\clockconstraint}, \bar{\variableconstraint}, \bar{\action}, \bar{\resettedclocks}, \bar{\varassignement}} \loc_{h+1}^\prime[ \bar{\autindex} ]$, of two distinct TA $\autindex,\bar{\autindex}$, which both set the value of variable $\variable$ (in a compatible manner), then the edge of the transitions is the same; that is, either they are $\action^{](}$ and $\bar{\action}^{](}$, or they are $\action^{)[}$ and $\bar{\action}^{)[}$.
\end{definition}

In the rest of the paper, only traces that are edge-consistent are considered.

Let $(\loc,$ $\variablevaluationfunction,$ $\clockvaluationfunction)$ be a configuration; we denote as  $c(\loc, \variablevaluationfunction, \clockvaluationfunction)$ 
the pair $(\cup_{1\leq \autindex \leq \tanumber} L(\loc[\autindex]), \variablevaluationfunction) \in \partSet(AP) \times \intNum^\variables$
of the atomic propositions and variable assignments that hold in the configuration $(\loc,$ $\variablevaluationfunction,$ $\clockvaluationfunction)$.
Let \tracesymbol\ be an edge-consistent trace 
{$
(\loc_0, \variablevaluationfunctionpar{0}, \clockvaluationfunctionpar{0}) 
\xrightarrow{\delta_0} 
(\loc'_0, \variablevaluationfunctionpar{0}', \clockvaluationfunctionpar{0}') 
\xrightarrow{\Lambda_0}
(\loc_1, \variablevaluationfunctionpar{1}, \clockvaluationfunctionpar{1}) 
\xrightarrow{\delta_1}  \ldots
$; we indicate by $\eventtime(e)$ the ``time'' of a symbol $e$ (where $e$ can be either $\delta$ or $\Lambda$), defined as follows:}
\begin{itemize}
	\item $\eventtime(\delta_0) = 0$;
	\item $\eventtime(\Lambda_h) = \eventtime(\delta_{h}) + \delta_{h}$ {for all $h\geq 0$};
	\item $\eventtime(\delta_h) = \eventtime(\Lambda_{h-1})$ {for all $h > 0$}.
\end{itemize}
Finally, let $w(\eta)$ be
the sequence $\Lambda_0 \delta_1 \Lambda_1 \delta_2 \dots$.

\begin{definition}\label{def:signal}
{Let \tracesymbol\ be an edge-consistent trace of a network $\netid$ of $\tanumber$ TA.
}
	The \emph{signal} \signalsymbol\ associated with \tracesymbol\ is the function $\signalsymbol:   \realNum_{\geq 0} \rightarrow \partSet(AP) \times \intNum^\variables $ such that:
	\begin{enumerate}
		\item\label{signal-time-0} $\signalsymbol(0)=c(\loc_0, \variablevaluationfunctionpar{0}, \clockvaluationfunctionpar{0})$;
		\item\label{signal-time-t} for all $\delta_h$ in $w(\tracesymbol)$, for all $r \in \realNum_{\geq 0}$ such that $\eventtime(\delta_h) < r < \eventtime(\delta_h)+\delta_h$
		then $\signalsymbol(r)=c(\loc_h, \variablevaluationfunctionpar{h}, \clockvaluationfunctionpar{h})$;
		\item\label{signal-discrete-t} for all $\Lambda_h$ in $w(\tracesymbol)$, $\signalsymbol(\eventtime(\Lambda_h))=(A, \variablevaluationfunction) \in \partSet(AP) \times \intNum^\variables$ where, for all $p \in AP$ and $n \in Int$:
		\begin{enumerate}
			\item \label{signal-discrete-t-atoms} $\proposition \in A$ if, for some $\action \in \actions$ and for some $1\leq \autindex \leq \tanumber$: 
			\begin{itemize}
				\item $\proposition \in L(\loc_h[\autindex])$ and $\Lambda_h[\autindex] \in \{\_, \action^{](}\}$ holds, or
				\item $\proposition \in L(\loc_{h+1}[\autindex])$ and $\Lambda_h[\autindex] = \action^{)[}$ holds
			\end{itemize}
			\item \label{signal-discrete-t-var-I} $\variablevaluationfunction(\variable) = \variablevaluationfunctionpar{h}(\variable)$ if one of the following conditions holds:
			\begin{itemize}
				\item there is no transition $\loc'_{h}[\autindex] \xrightarrow{\clockconstraint, \variableconstraint, \action, \resettedclocks, \varassignement} \loc_{h+1}[\autindex]$ compatible with the configuration change and such that $\variable\in U(\mu)$; 
				\item 
				 there is $1\leq \autindex\leq \tanumber$ and a transition $\loc'_{h}[\autindex] \xrightarrow{\clockconstraint, \variableconstraint, \action, \resettedclocks, \varassignement} \loc_{h+1}[\autindex]$---compatible with the configuration change---such that $\Lambda_h[\autindex] = \action^{](}$ and $\variable\in U(\mu)$.
			\end{itemize} 
			\item \label{signal-discrete-t-var-II} $\variablevaluationfunction(\variable) = \variablevaluationfunctionpar{h+1}(\variable)$ if there is $1\leq \autindex\leq
			\tanumber$ and a transition $\loc'_{h}[\autindex] \xrightarrow{\clockconstraint, \variableconstraint, \action, \resettedclocks, \varassignement} \loc_{h+1}[\autindex]$---compatible with the configuration change---such that $\Lambda_h[\autindex] = \action^{)[}$ and $\variable\in U(\mu)$ hold.
			%

		\end{enumerate}
		
	\end{enumerate}
\end{definition}

}

{When networks of TA are considered, the event symbols labeling the transitions are used to synchronize automata. 
}
Two (or more) different TA can take their transitions at the same time by labeling them with the same synchronization channel, and using the actions to describe the type of synchronization desired.
{Every event symbol $\action \in Act$ is associated with one communication channel, which can be identified with the event symbol itself---i.e., channel $\action$.
}
%
The first type of synchronization is one-to-one synchronization.
A transition labeled with one-to-one send $\alpha{!}$, for some channel $\alpha$, can only be fired if at the same moment in time, another TA takes a transition labeled with the one-to-one receive $\alpha{?}$.
The second type of synchronization available is termed `broadcast' synchronization. 
Like one-to-one synchronization, for a given channel $\alpha$ there can only be one active transition with the broadcast-send $\alpha\#$, however the difference is that there can be 0, 1, or multiple automata that sync using broadcast-receive $\alpha@$ at once.
In addition, each automaton is \emph{required} to perform a broadcast-receive if it is able to, meaning that there exists a transition $t$ such that $t_{-}$ is the currently active state, and all guards of the transition are satisfied.
{The details of the semantics of synchronizations can be found in \cite{tack20}.}

\subsection{Constraint LTL over clocks}
\label{sec:cltloc}

Constraint LTL over clocks (\cltloc{}) is an extension of LTL where
formulas are defined over atomic propositions and clocks.
A clock is a variable over \(\mathbb{R}_{\geq 0}\) whose value changes between positions in a { \cltloc{} model} to represent the passage of time.
In addition, \cltloc\ has been extended to support expressions over arithmetical variables~\cite{marconi16}.

A formula in \cltloc\ consists of atomic propositions, clock formulas, and
formulas over integer variables, which are combined using the standard LTL
operators of \(\mathcal{X}\) (next) and \(\mathcal{U}\) (until), as well as the
derived operators \(\mathcal{G}\) (globally), \(\mathcal{F}\) (future), and
\(\mathcal{R}\) (release). A clock formula compares the value of the clock to a
given natural number, for instance \(x > 7\). A variable formula, on the other
hand, can compare not only individual variables but also arithmetic combinations
of variables. An example would be the expression \(b + c = 7\); \(b,c \in \Int\).
{\cltloc{} uses a special version of the $\mathcal{X}$ operator that can be applied to  variables in $\Int$.
A valid formula is, for instance, $\mathcal{X}(n) = n + 1$.\footnote{{It is easy to see that TA with variables and \cltloc{} as defined in Sec.~\ref{timed-automata} and \ref{sec:cltloc} are undecidable, unless suitable restrictions are introduced. In this paper we consider variables with finite domains.}}
}

Let \(X\) be a finite set of clocks and \(\Int\) be a finite set of integer variables.
\cltloc\ formulas are defined as follows:
\[
\begin{aligned}
\phi := & \pi \BitOr x \sim c \BitOr \exp_{1} \sim \exp_{2} \BitOr \mathcal{X}(n) \sim \exp \BitOr \\
        & \phi \land \phi \BitOr \neg \phi \BitOr \mathcal{X}\phi \BitOr \phi \mathcal{U} \phi
\end{aligned}
\]
where \(\pi \in AP\), \(x \in X\), \(c \in \mathbb{N}\), \(n \in \Int\),
$\sim \in \{<,=\} $ and \(\exp\) are arithmetic formulas over integer variables
and integers (defined in Section~\ref{timed-automata}).

Like in TA, clocks are special dense variables over \(\mathbb{R}_{\geq 0}\) that ``progress'' between different positions along a {\cltloc{} model}:  each clock must either increment between two adjacent time positions, or it must be reset.
We introduce \(\delta: \mathbb{N} \rightarrow \mathbb{R}_{>0}\), which measures the amount of time that elapses between two adjacent time positions.
For a given clock valuation \(\sigma: \mathbb{N} \times X \rightarrow \mathbb{R}_{\geq 0}\), each clock \(x \in X\) must either obey the equivalence \(\sigma(l,x) + \delta(l) = \sigma(l+1,x)\), or is reset, i.e.\ \(\sigma(l+1,x) = 0\) holds.
%
We also define variables via the assignment function $\iota : \mathbb{N} \times \Int \rightarrow \mathbb{Z}$ that assigns a value to each variable $n \in \Int$ at every time position in $\mathbb{N}$.
The arithmetical expressions $\exp$ can now be evaluated at a time position $l$ by replacing every occurrence of an integer variable $n$ with $\iota(l,n)$.

For the sake of space, we do not provide in this paper the full formalization of the semantics of \cltloc{} and we refer the reader to \cite{tack20}, instead.

\subsection{TACK \cltloc{}-based Translation}
\label{prelim-tack}

The TACK~\cite{tack20} tool allows users to perform the formal verification of TA against properties expressed in Metric Interval Temporal Logic (MITL, \cite{AFH96}).
To this end, TACK takes as input a TA network $\mathcal{N}$ and a MITL formula $\phi$ to be checked, transforms both of them into suitable \cltloc{} formulas, and uses the Zot tool, which supports the formal verification of \cltloc{} formulas through a Bounded Satisfiability Checking approach \cite{PRB20}, to automatically verify whether property $\phi$ holds for $\mathcal{N}$ or not.
In the rest of this section we provide an overview of the TA-to-\cltloc{} translation performed by TACK, which is the subject of the improvements presented in Section \ref{sec:ImprovEnc}.
Notice that, instead, the encoding of MITL properties is done following the approach defined in \cite{bersani15}, which was also applied in \cite{tack20}.

In \cite{tack20}, the \cltloc{} formula constructed from a given network of TA represents the evolution of the configuration (i.e., an execution) of the network over the continuous time.
As mentioned in Section \ref{timed-automata}, we adopt a semantics of TA based on signals, and a configuration captures the value of all the clocks, variables and current locations of the TA in the network in a specific time instant.
The key aspect of the encoding described in \cite{tack20} and \cite{bersani15} is that every time position of a {model}  satisfying the \cltloc{} formula is representative for a nonempty interval of $\mathbb{R}_{\geq 0}$.
{In addition, in \cite{tack20}, every \cltloc{} model satisfying the formula encoding a network $\mathcal{N}$ represents an execution of the network, i.e., a trace $\eta$ of $\mathcal{N}$. Hence, it is an exact representation of signal $\signalsymbol$.
}

The most relevant part of the \cltloc{} formula is the encoding of the firing of transitions and the possible synchronization among them, which precisely capture the dynamics of the variables, clocks and location changes between any two adjacent time intervals over the signals.

At every time position of the \cltloc{} model, function $p[i]$ (with $i \in [1,N]$) represents the placing of $\mathcal{A}_i$, i.e., the active location in the corresponding interval of the execution of the TA $\mathcal{A}_i$ of network $\mathcal{N}$, and function $t[i]$ represents the transition that will be taken at the end of the interval (as depicted in Figure \ref{fig:differentsemantics} for a single TA).
Each function is syntactic sugar for a finite set of atomic propositions, \repMB{one for each possible value of the function, that are constrained so that only one may evaluate to true in each time position.}{which encode the value of a single variable---e.g., the current location of an automaton---belonging to finite set of values.}
When a transition is taken, proposition $edge_{i}^{RC}$ represents the edge ($]($ or $)[$) with which it is taken---i.e., if the current interval of the signal associated with the $i$-th automaton is left- or right-closed. TA clocks and variables can be represented directly as \cltloc\ clocks and variables.


\begin{table*}
  \centering
  \aboverulesep=0ex
  \belowrulesep=0ex
  \renewcommand{\arraystretch}{1.2}
  \caption{Snippet of TACK encoding of a TA in CLTLoc}
  \label{tack-encoding}
  \begin{tabular}{c|c|c}
    \toprule
    \multicolumn{3}{c}{\(\varphi_{6} := \underset{q \in \mathcal{Q}_{i}}{\underset{i \in [1,N]}{\bigwedge}} \bigg( \Big( p[i] = q \land t[i] = \sharp \Big) \rightarrow \mathcal{X} \Big( \Inv(q) \land r_{1}(\Inv(q)) \Big) \bigg) \)} \\
    \midrule
    \multicolumn{3}{c}{\( \varphi_{7} := \underset{t \in T_{i}}{\underset{i \in [1,N]}{\bigwedge}} t[i] = t \rightarrow \Big( p[i] = t_{-} \land \mathcal{X}(p[i] = t_{+}) \land \varphi_{\gamma_{c}} \land \varphi_{\gamma_{v}} \land \varphi_{\alpha_{c}} \land \varphi_{\alpha_{v}}  \land \varphi_{edge}(t_{-}, t_{+}, i) \Big) \)} \\
\bottomrule
  \end{tabular}
\end{table*}

Table~\ref{tack-encoding} contains a snippet of the formulas used to encode executions of networks of TA into CLTLoc.
The formulas use propositions $p[i]$, $t[i]$ and $edge_{i}^{RC}$ introduced above.
Notice that not every TA needs to transition at each time position (for example, at a given point in the execution $\mathcal{A}_1$ might change location, whereas $\mathcal{A}_2$ does not take any transition).
Hence, the encoding introduces a \emph{null transition} symbol \(\sharp\) to represent the situation in which no transition is taken.
So, function \(t[i]\) is equal to either a transition or the symbol \(\sharp\) (see Figure \ref{fig:differentsemantics}).
If transition \(t\) is active in a given position $l$ of an execution for automaton $\mathcal{A}_i$, then the TA is in location \(t_{-}\) in position $l$, and in location \(t_{+}\) at the next one (i.e., in interval $l+1$).



Formula $\varphi_{6}$ defines the semantics for the null transition.
If TA $\mathcal{A}_i$ performs a null transition, the state invariant must hold both before and after clock resets are applied.
Function $r_{1}$ replaces the value of any reset clock with $0$, thus capturing the post-reset value of any clock used in the invariant.

Formula $\varphi_{7}$ encodes the discrete transitions.
Each must respect the guards and assignments of the transitions, the TA must currently be in the source location of the transition, and must be in the destination location in the following position.
Formulas $\varphi_{\gamma_c}$, $\varphi_{\gamma_v}$, $\varphi_{\alpha_c}$, $\varphi_{\alpha_v}$ capture the guards and assignments associated with the transition.
$\varphi_{edge}$ encodes the two possible edge configurations, right- and left-closed, and ensures that the invariants are satisfied
depending on the edge type.

The encoding includes many other formulas, for example to define the initial values of variables and clocks, or the sufficient conditions for a transition to be taken, but they are not shown here for the sake of brevity.
For the same reason we do not show here the \cltloc{} formulas capturing the synchronization mechanisms among the TA of a network and those related to various liveness constraints supported by the TACK tool.
Interested readers can refer to \cite{tack20} for further details.

\section{Improved Encoding}
\label{sec:ImprovEnc}

In the TACK tool, the \cltloc{} formulas produced through the encoding presented in Section \ref{prelim-tack} are fed to the Zot formal verification tool, which in turn suitably translates them into the input logics (and in particular BitVector logic) of Satisfiability Modulo Theories (SMT) solvers.
In this section we present a novel method---named \newTACK{}---for encoding executions of networks of TA into the logics supported by SMT solvers.
The method skips the intermediate \cltloc{} representation to directly produce formulas of BitVector logic to be fed to SMT solvers.
This direct translation allows us to make several optimizations not possible in \cltloc.
As before, the MITL property will continue to be converted first into \cltloc\ before being transformed into BitVector logic by TACK through Zot.
We use a consistent naming convention for the atomic propositions to ensure that the two BitVector encodings (the one for TA and the one for MITL formulas)
can be safely combined to produce the final SMT output.
This section first describes the various terms that make up our TA network, and discusses how they are encoded into BitVector logic.
Then, it overviews of the constraints, defined over the terms previously defined, that capture the TA semantics.
Finally, it provides an argument for the correctness of the new encoding, and highlights the improvements made over the original encoding.
For ease of reading we will refer to the old encoding as \oldTACK\ when contrasting it with \newTACK{}.
For the sake of space, this paper does not present the full \newTACK{} encoding; interested readers can refer to \cite{smith20}
for further details.

\subsection{BitVector-based representation of terms}
\label{sec:terms}

Our novel encoding (\newTACK{}) is based on the idea of directly representing the terms of the TA into BitVectors.
Since we are using a \emph{bounded} verification approach, our goal is to represent the terms over a finite number $k+2$ of discrete positions.
Using BitVector logic, we can group logically connected propositions into a BitVector, which results in a more compact encoding and can grant significant speedups on operations performed over every element of the vector.

\subsubsection{Transitions}\label{encoding-transitions}

Before describing the BitVector terms for the transitions, we must make one key change to our set of transitions.
For reasons to be discussed we wish to represent the \emph{null transition} (when a TA does not transition between time positions) not as the separate entity \(\sharp\), but rather as a set of
$|Q_{i}|$ transitions, one for each location $q \in Q_{i}$. 
\[\underset{i \in [1,N]}{\forall}\ \underset{q \in Q_{i}}{\forall}\ t_{null_q} := {<}q, q, \tau, \varnothing, \varnothing, \varnothing, \varnothing {>}\]
These null transitions have the same source and destination location, and no
constraints or assignments. We can now refer to the set of all transitions as
\(\mathcal{T}\), defined as
\(\mathcal{T}_{i} = \underset{q \in Q_{i}}{\bigcup}\{ t_{null_{q}}\} \cup T_{i}\)
for each TA \(\mathcal{A}_{i}\). As before \(\mathcal{T}\) is the union of the
\(\mathcal{T}_{i}\) sets. The motivation for this redefinition will become clear
when we discuss the encoding of the active locations of the TA.

To encode $\mathcal{T}_i$,
we adopt a similar approach as the one used in \oldTACK.
%
Rather than store each transition as a separate BitVector, 
since only one transition is active at a time in automaton $\mathcal{A}_i$, we store the currently active transition as a binary number over \(\lceil\log_2 |\mathcal{T}_i|\rceil\) bits.
Therefore, we create \(\lceil\log_2 |\mathcal{T}_i|\rceil\) BitVectors $tb_{i,0}, tb_{i,1}, \ldots, tb_{i, \lceil \log_2 |\mathcal{T}_i| \rceil -1}$ of length \(k+2\), \addMB{each one representing a single bit of a  numeric identifier that encodes the transitions in $\mathcal{T}_i$}, 
\addMB{i.e., the $l$-th bit of vector $tb_{i,j}$ is the $j$-th digit (weight $2^j$) of the binary number which} 
indicates the active transition of $\mathcal{A}_i$ \repMB{over time}{at the time position $l$} (see \cite{PRB20} for details about the principles behind bounded BitVector-based encodings).
\addMB{For the sake of convenience, to easily identify the time position in which a transition is taken, we associate every transition $t \in \mathcal{T}_{i}$ with a BitVector, whose $l$-th bit has the value of 1 if $t$ is active at position $l$ (firing occurs at $l+1$).
The vector is determined using bit-wise logical operations over $tb_{i,0}, tb_{i,1}, \ldots, tb_{i, \lceil \log_2 |\mathcal{T}_i| \rceil -1}$.
}
For example, suppose that a transition $t \in \mathcal{T}_{i}$ is active at positions $1$ and $3$ of a bounded sequence such that $k = 8$.
That information is represented by BitVector $\overleftarrow{0000001010}$ of length $k+2$.
Now, consider \(\lceil\log_2 |\mathcal{T}_i|\rceil = 6\) and a transition \(t \in \mathcal{T}_i\) whose identifier is $5$. 
Since the binary representation of $5$ is $000101$, we express the CNF representation (maxterm) for the value 5 with BitVector variables $tb_{i,j}$ and construct formula
\[
\BitNeg \overleftarrow{tb}_{i,5}\ \BitAnd\ 
\BitNeg \overleftarrow{tb}_{i,4}\ \BitAnd\ 
\BitNeg \overleftarrow{tb}_{i, 3}\ \BitAnd\
\overleftarrow{tb}_{i, 2}\ \BitAnd\ 
\BitNeg \overleftarrow{tb}_{i, 1}\ \BitAnd\
\overleftarrow{tb}_{i, 0}.\]
\addMB{that defines a BitVector of length $k+2$ such that the $l$-th bit is 1 if transition $t=5$ is active at time position $l$}



\begin{table}
{\small
\centering
\begin{tabular}{c | c}
Transition & Alias \\
\midrule
\(\mathcal{T}_{i}[0]\) & \(\ \BitNeg\overleftarrow{tb_{i,\lceil \log_2 \mathcal{T}_i \rceil -1}} \BitAnd \
\ldots \BitAnd \ \BitNeg\overleftarrow{tb_{i,1}} \BitAnd \ \BitNeg\overleftarrow{tb_{i,0}} \) \\
\(\mathcal{T}_{i}[1]\) & \(\ \BitNeg\overleftarrow{tb_{i,\lceil \log_2 \mathcal{T}_i \rceil -1}} \BitAnd \
\ldots \BitAnd \ \BitNeg\overleftarrow{tb_{i,1}} \BitAnd \overleftarrow{tb_{i,0}} \) \\
\(\mathcal{T}_{i}[2]\) & \(\ \BitNeg\overleftarrow{tb_{i,\lceil \log_2 \mathcal{T}_i \rceil -1}} \BitAnd \
\ldots \BitAnd \overleftarrow{tb_{i,1}} \BitAnd \ \BitNeg\overleftarrow{tb_{i,0}} \) \\
\rotatebox{90}{\(\ldots\)} & \rotatebox{90}{\(\ldots\)} \\
\(\mathcal{T}_{i}[|\mathcal{T}_{i}|]\) & \(\overleftarrow{tb_{i,\lceil \log_2 \mathcal{T}_i \rceil -1}} \BitAnd
 \ldots \BitAnd (\sim\overleftarrow{tb_{i,1}}) \BitAnd (\sim\overleftarrow{tb_{i,0}}) \) \\

\end{tabular}
}
\caption{Construction of the Transition Aliases}\label{t-aliases}
\end{table}

We use expression such as the one above to define aliases for the \(|\mathcal{T}_{i}|\) transitions of TA $\mathcal{A}_i$, as shown in Table~\ref{t-aliases}, such that each transition is identified by means of a unique alias, i.e., each transition is encoded as a unique combination of the \(tb_{i,j}\) vectors.
We indicate the alias for a transition $t$ whose identifier is $h$ as $\overleftarrow{t}$ or $\overleftarrow{\mathcal{T}_i[h]}$, depending on the case.
Because the exact value of \(|\mathcal{T}_{i}|\) is variable, for the last transition in the table we use the symbol \(\sim\) to signal that whether or not the BitVector is negated depends on the exact value of \(|\mathcal{T}_{i}|\).

Consider now transition edges. 
We introduce a BitVector \(\overleftarrow{edge_{i}^{RC}},\ i \in [1,N]\) of length $k+2$ for each TA in the network.
When a bit is set to \(1\) (resp., $0$), it signifies that the active transition for the TA at that time position is right-closed (resp., left-closed).

\subsubsection{Location}\label{encoding-states}

\addMB{For every location $q \in Q_i$, we introduce an alias defining a vector of $k+2$ positions that indicate if the current location of automaton $\mathcal{A}_i$ is $q$.}
Since 
the active location of $\mathcal{A}_i$ is the source location  \(t_{-}\) of the active transition,
we define location $\overleftarrow{q}$ as the bit-wise disjunction of all the transitions whose source is $q$.
\[\underset{q \in Q_i}{\forall}\ \ \overleftarrow{q} := \underset{t \in \mathcal{T}_i|t_{-} = q}{\BitOr}\overleftarrow{t}\]
This is made possible by our addition of $|Q_{i}|$ null transitions, 
\addMB{one for each location.}
This was not possible in TACK's \cltloc\ encoding because of the use of a single null transition per automaton.
When the \cltloc\ null transition is active, it is not possible to determine the active locations without referring to variable $p[i]$.

\subsubsection{Variables}\label{encoding-variables}

Unlike location and transitions, the possible values of a bounded integer variable are not unrelated objects in a set, but 
their value 
must respect the operations of addition and subtraction.
For each variable \(n \in \Int\) we 
construct a bit representation \(\overleftarrow{vb_{n,j}}\), where each BitVector has length \(k+2\).
The values are encoded in twos complement notation, and the number of BitVectors is chosen so that the vectors are capable of representing the entire range of values for the given bounded integer variable.
We will define \(\lambda(n)\) as the number of bits needed for each variable \(n\).

To refer to the complete value of a variable at a particular time position, rather than a particular bit of the variable, 
we make use of the \emph{extract} and \emph{concat} BitVector logic operators to define a second set of BitVectors \(\overleftarrow{var_{n}(l)}\) of \(\lambda(n)\) bits, defined over the vectors $\overleftarrow{vb_{n,j}}$ that represents the value of variable \(n\) at time position \(l\), with \(0 \leq l \leq k+1\).

\subsubsection{Clocks}\label{encoding-clocks}

Our encoding of the clocks does not differ from \oldTACK. Each clock
\(x \in X\) is defined as a function \(x(l)\) that takes an integer
argument $l$ corresponding to a time position and returns a real number representing the value of $x$ at position $l$.

\subsubsection{Complete Encoding of Terms}\label{complete-encoding}

A valid trace of the network consists of assigning values to the terms described above.
To build valid traces, we define a number of constraints that make use of two helper terms, \(\delta\) and \(\overleftarrow{loop}\). 
The first one represents the amount of time that passes between two adjacent time positions (i.e., the length of the corresponding interval), and must be a positive real number.
The second, the term \(\overleftarrow{loop}\), has a value equal to the index of the first time position in the loop portion of the trace.
From these we can represent any valid lasso-shaped trace of the network of length $k+2$, as typically done in bounded verification approaches (see also \cite{PRB20}).
{In particular, a constraint limits the position of the loop to be a positive value bounded by $k$.
The constants 0 and $k$ are encoded using BitVectors of length $k+2$.
For instance, the value $4$ over 5 bits would be written as $\overleftarrow{4}_{[5]}$ (in this case we use the subscript to make the length of the BitVector explicit) and expands to $00100$.
Since BitVector logic supports arithmetic, the relation $<$ can be applied to express the bounds for term \(\overleftarrow{loop}\) as follows:
\[
\overleftarrow{0} < \overleftarrow{loop} < \overleftarrow{k}.
\]
}

In addition, we introduce aliases to more easily refer to the transitions and locations individually, and  to the value of a variable at a particular time position.

\subsection{Constraints}\label{constraints}

The terms introduced in Section \ref{sec:terms} allow us to describe lasso-shaped traces of networks of TA, but we need to introduce suitable restrictions to avoid capturing traces that do not respect the signal-based semantics of TA.
These restrictions take the form of clock guards on a transition, location invariants that prevent a TA from staying in a location indefinitely, clock progression constraints, and so on.
We formalize these constraints in BitVector logic for the SMT solver to use when performing the Bounded Model Checking of TA.
For brevity, in this paper we dot not present the full set of constraints; Table~\ref{table:constraints-trans} shows some significant formulas, which are explained in the rest of this section to illustrate how the terms introduced above impact on the new \newTACK{} encoding.
Further details can be found in \cite{smith20}.

{Formula $\phi_5$} ensures that the active location of a TA correctly reflects the transition being taken.
It asserts that when a transition is taken at position \(l\), the destination location is active at position \(l{+}1\) ($\overleftarrow{t}^{[k:0]}$ indicates that we are considering the bits of BitVector $\overleftarrow{t}$ in range $[0,k]$).
Because the location BitVectors are just aliases defined over the transition BitVectors (see Section \ref{encoding-states}), we do not need to explicitly constrain the TA to be in location \(t_{-}\) at time position \(l\), since this is true by definition.

\begin{table*}
\caption{Snippet of transition constraints for a network of TA. Terms $\sigma$ and $\zeta$ are based on grammars presented in Sec.~\ref{sec:Prelim}. }
\centering
\begin{tabular}{c @{\hskip 1cm} c}
\midrule
  \(\phi_5 := \underset{t \in \mathcal{T}}{\bigwedge} (\overleftarrow{t}^{[k:0]} \rightarrow \overleftarrow{{t_+}}^{[k+1:1]})\)
  &
  \(\phi_9 := \underset{t \in T}{\bigwedge}\ \underset{l \in [0,k]}{\bigwedge} \overleftarrow{t}^{[l]} \rightarrow \sigma_{\delta}(l,t_{\gamma_{c}}) \)
   \\
\midrule
{\(\phi_{10} := \)\(\underset{t \in T}{\bigwedge}\ \underset{l \in [0,k]}{\bigwedge} \overleftarrow{t}^{[l]} \rightarrow \mu(l,t_{\gamma_{v}}) \)}
&
{\(\phi_{11} := \)\(\underset{t \in T}{\bigwedge}\ \underset{x \in t_{a_c}}{\bigwedge}\ \underset{l \in [0,k]}{\bigwedge} \overleftarrow{t}^{[l]} \rightarrow x(l{+}1) = 0\)} \\
\midrule
\multicolumn{2}{c}{ \(\phi_{12} := \underset{t \in T}{\bigwedge}\ \underset{n,\exp \in t_{a_v}}{\bigwedge}\ \underset{l \in [0,k]}{\bigwedge} \overleftarrow{t}^{[l]} \rightarrow \big(\overleftarrow{var_{n}(l{+}1)} = \overleftarrow{\zeta(l,n,\exp)}\big) \)}
\\
\midrule
\multicolumn{2}{c}{  $\phi_{13} := 
                       \underset{t \in T_{i}}{\underset{i \in [1,N]}{\bigwedge}}\ \underset{l \in [0,k]}{\bigwedge} \overleftarrow{t}^{[l]} \rightarrow 
                       \begin{array}{l}
                       \biggl(\sigma_{\delta}(l, \Inv(t_-)) \land \sigma_{w}(l{+}1, \Inv(t_+)) \land (\overleftarrow{edge_{i}^{RC}}^{[l]} = \overleftarrow{1})\biggr) \ \lor
                       \\
                       \biggl(\sigma_{w\delta}(l, \Inv(t_-)) \land \sigma(l{+}1, \Inv(t_+)) \land (\overleftarrow{edge_{i}^{RC}}^{[l]} = \overleftarrow{0})\biggr)
                       \end{array}
                       $} \\
\midrule
\( \sigma(l,\gamma_{c}) :=  x(l) \sim c\ |\ \sigma(l,\gamma_{c}') \land \sigma(l,\gamma_{c}'') \) &
  \( \sigma_{\delta}(l,\gamma_{c}) := x(l) + \delta(l) \sim c\ |\ \sigma_{\delta}(l,\gamma_{c}') \land \sigma_{\delta}(l,\gamma_{c}'') \) \\
\midrule
\( \sigma_w(l,\gamma_{c}) :=  x(l) \sim_{w} c\ |\ \sigma_{w}(l,\gamma_{c}') \land \sigma_{w}(l,\gamma_{c}'') \)  &
\( \sigma_{w\delta}(l,\gamma_{c}) :=  x(l) + \delta(l) \sim_{w} c\ |\ \sigma_{w\delta}(l,\gamma_{c}') \land \sigma_{w\delta}(l,\gamma_{c}'') \)
 \\
\midrule
\multicolumn{2}{c}{
{\( \mu(l,\gamma_{v}) := \overleftarrow{var_{n}(l)} \sim \overleftarrow{c}\ |\ \overleftarrow{var_{n}(l)} \sim \overleftarrow{var_{n'}(l)}\ |\ \neg \mu(l,\gamma_{v}')\ |\ \mu(l,\gamma_{v}') \land \mu(l,\gamma_{v}'') \)}} \\
\midrule
\multicolumn{2}{c}{ \( \zeta(l,n,\exp) :=  \overleftarrow{var_{n}(l)}\ |\ \overleftarrow{c}\ |\  \zeta(l,n,\exp') + \zeta(l,n,\exp'')\ |\  \zeta(l,n,\exp') - \zeta(l,n,\exp'') \)} \\
\bottomrule
\\
\end{tabular}
\label{table:constraints-trans}
\end{table*}


Each transition can have multiple guards, which consist of two types, clock guards and variable guards.
Formula \(\phi_9\) asserts that, for every clock guard, its associated transition being active at time position \(l\) implies that at the instance of transition, the relationship \(\sim\) holds between the clock value and the value $c$.
Recall that if a transition is active at position \(l\), the transition occurs in the instant 
corresponding to the position \(l+1\),
where clock $x$ does not have the value \(x(l)\), but rather \(x(l) + \delta(l)\). 
Note that we cannot simply use \(x(l+1)\) as the value of the clock in \(\phi_9\), because it is possible that the transition 
can reset $x$ at $l+1$, with $x(l+1)=0$ being the post-transition value.
The guard only sees the pre-transition value of the clock, thus we must explicitly add \(\delta(l)\) to $x(l)$.
{The term $\sigma_{\delta}(l,t_{\gamma_c})$ 
is the encoding of clock constraint $t_{\gamma_c}$ expressed at position $l$ and considering the time delay between position $l$ and $l+1$ stored in $\delta$.}

{Formula $\phi_{10}$ captures the same semantics for
variable guards, asserting that an active transition implies that the variable guard is true at that time position. 
Because variables, unlike clocks, do not progress with time, it is sufficient to simply use the value \(var_{n}(l)\)
to determine if the guard is satisfied. The function $\mu$ is used to encode the variable constraint grammar. 
If the form
$\overleftarrow{var_{n}(l)} \sim \overleftarrow{var_{n'}(l)}$
is used and $\lambda(n') < \lambda(n)$, then $\overleftarrow{var_{n'}(l)}$ is implicitly sign-extended to a
length of $\lambda(n)$ bits (conversely, it is truncated). 
}

{Formula $\phi_{11}$ models clock assignments, that are more straightforward than the clock guards.
It is enough to require that if a transition is taken at time position \(l\), then in the following time position the clock is reset.}

Formula $\phi_{12}$ captures the semantics of variable assignments.
Variable assignments can refer to both constant values and the values of other variables, and they may combine them using the operators \(\{+,-\}\).
To implement this in our BitVector logic, we require that if any variable \(n'\) appears in the assignment expression of variable \(n\), then $\lambda(n') \leq \lambda(n)$ holds. 
We can then cast all constants and variables to BitVectors of length \(\lambda(n)\), sign-extending shorter values to a length of $\lambda(n)$ bits if necessary.
This allows us to use the standard BitVector addition and subtraction operators to compute the final value, which is assigned to \(v\) at time position \(l{+}1\).
{The term $\zeta(l,n,\exp)$ encodes the expression $\exp$ with the values of arithmetical variables at position $l$}.

Formula $\phi_{13}$ captures the semantics of location invariants.
Although invariants are location-specific, not transition-specific, since locations are defined by the active transitions, it is sufficient to ensure that at the moment of transition both the source and destination invariants are satisfied, taking into account the value of $\overleftarrow{edge_{i}^{RC}}$.
Since all invariants are convex, if the invariant is satisfied at moment the TA enters the location and at the moment it leaves, it is satisfied at all positions in the interval between them.
The occurrence of a transition at position $l$ implies one of two statements, one for each possible value of $\overleftarrow{edge_{i}^{RC}}$.
{In both the statements, the invariants of the source location are evaluated by considering the pre-transition clock values at position $l+1$, i.e., $x(l)+\delta(l)$, hence using the terms $\sigma_\delta$ and $\sigma_{w\delta}$, as the clock resets have not happened yet. 
Conversely, the invariants of the destination location are evaluated by considering post-transition clock values at position $l+1$, hence using the terms $\sigma$ and $\sigma_{w}$.
In addition, the invariant of the location (either $t_-$ or $t_+$, depending on $edge_i^{RC}$) that is not the current location of the automaton at the time instant in which a transition occurs, i.e., whose signal has an open-ended edge transition, are evaluated with the weak satisfaction relation $\sim_w$ (the interested reader can find the definition in \cite{tack20}).}

\color{black}
The complete \newTACK{} encoding includes, in addition to the formulas of Table~\ref{table:constraints-trans} (which we can conjoin in a single formula, $\phi_{trans}$), formulas that govern the initialization and progression of the TA ($\phi_{init}$), formulas that capture the semantics of synchronizations ($\phi_{sync}$), and formulas that guarantee the correctness of the lasso-shaped traces ($\phi_{loop}$).
Overall, the encoding of the semantics of a network of TA $\mathcal{N}$ is given by the following formula (we refer to \cite{smith20} for details):
\[\phi_{\mathcal{N}} := \phi_{init} \land \phi_{trans} \land \phi_{sync} \land \phi_{loop}\]

\subsection{Equivalence and Improvements}\label{equivalence}


In this section we outline an argument showing that the \newTACK{} encoding given by formula $\phi_{\mathcal{N}}$ of Section \ref{constraints} is a correct and complete representation of all lasso-shaped, non-Zeno runs of length \(k+2\) of network $\mathcal{N}$.
More precisely, we briefly compare the \newTACK{} and \oldTACK{} encodings and show that they capture the same constraints.
Hence we conclude that the \newTACK{} encoding is sound and complete, since \oldTACK{} has been proved to be so in \cite{tack20}.
We also highlight the points in which \newTACK{} improves on \oldTACK{}.

Both the \oldTACK{} and the \newTACK{} encodings constrain the clocks, variables, and TA to their respective initial values and locations at time position 0.
For variables and clocks these constraints are identical, as both assign the desired value at time position 0.
For locations \newTACK{} uses the \(\overleftarrow{{q}}\) aliases to require that the TA begins in the initial location, despite not having location BitVectors.
Because the location alias is only true when one of the transitions whose source is that location is true (including the location-specific null transitions), the constraint is valid.
Function \(\delta(l)\) ensures that all clocks progress at the same rate, while clock resets and variable assignments are only allowed if one of the corresponding transitions are active.
As for the transitions, although we have broken up \(\varphi_{7}\) (see Table~\ref{tack-encoding}) into several pieces (some of which are shown in Table~\ref{table:constraints-trans}), the functionality remains the same. We ensure that in order for a
transition to be valid, its destination location must be active in the next time
position, the clock and variable constraints must be satisfied, all assignments
must be enforced, and the invariants of the source and destination location must be
true at the moment of transition.
Like \oldTACK{}, \newTACK{} allows that at the moment of transition, only one of the two invariants must be satisfied, using the concept of weak satisfaction to formalize this relaxation.
Similarly, the original \oldTACK{} encoding contains three constraints that assert that the values of the active locations, as well as the values of the variables and clocks, can only be changed if there is an active transition that modifies them.
For locations, this is accomplished with $\phi_{5}$, which requires that the active location in the following position be equal to the value of the destination location of the active transition.
Unlike in the original encoding, we have one null transition for each position, so we do not need to consider the null transitions as a special case.
Therefore for the location to change, there must be a non-null transition to enable the location change.
A pair of formulas, $\phi_{8}$ and $\phi_{7}$, not shown in this paper for brevity, assert that when no transition explicitly changes the value of a variable or resets a clock, their values remain the same.
Our new encoding also respects the same loop constraints as \oldTACK\, including the 
clock constraints
necessary to represent all possible lasso-shaped traces.

As shown in Section \ref{sec:Exp}, the new \newTACK{} encoding in many cases provides significant benefits in terms of efficiency of the verification procedure.
In addition, \newTACK{} introduces various improvements over \oldTACK{} concerning the range of TA features captured.
\oldTACK{} contains a limitation regarding integer variables: because they are represented as elements of a set, \oldTACK{} can only test them for equality.
This means that constraints of the form $n \sim c$ or $n \sim n'$, where $\sim \not\in \{=\}$ are not supported.
\newTACK{} correctly represents the values of the integer variables using a twos-complement encoding, and therefore can support the full grammar of variable guards and assignments.
The implementation of the \newTACK{} encoding in the TACK tool has also fixed some issues that were present in the old implementation of TACK, and it has allowed us to complete the set of features supported by the tool.
In particular, support for broadcast synchronization primitives in the old version of the TACK tool was faulty, and it has now been fixed in the implementation of \newTACK{}, as shown by our experimental results.
%
%
Finally, support for right-closed intervals, left-closed intervals, and arbitrary combinations thereof was not complete in the old implementation of TACK (only right-closed intervals were fully supported); \newTACK{}, instead, fully supports all types of intervals.

\section{Experimental Results}
\label{sec:Exp}

In this section we present the results of several experimental evaluations of the new \newTACK{} encoding compared with \oldTACK.
These tests cover several different benchmarks commonly used to evaluate formal verification techniques.
For both \oldTACK{} and \newTACK{}, strong transition liveness (see \cite{tack20}) was used in all of the tests, and all edges were constrained to be right-closed.
These were the settings used to benchmark the original TACK application, and they remain the default settings for the tool.
In all of the following tests, the measured time is the combined time taken by both the TACK program to parse the problem and convert it to SMT form and for the underlying Z3 solver \cite{de2008z3} to decide the satisfiability of the SMT problem.
In practice, the TACK translation always took less than a second.
For every test, the evaluation proceeded in several rounds, each with a larger length of traces considered by TACK.

All tests were performed running the Z3 SMT solver version 4.8.8 on a server equipped with an AMD EPYC 7551 CPU (2.5 GHz) with 2 32-core sockets, 500 GB of RAM and Debian Linux (version 4.19).
Although our tests were run on a large server with (at the time of writing!) an unusually high amount of both processors and RAM, the Z3 solver is a single-threaded application, and typically uses less than a gigabyte of RAM while running.
Therefore, 
very similar results could be obtained on a machine with more reasonable resources.
To reduce instabilities in the solver and to present a clearer comparison between the encodings, we used Z3's built-in `parallel-or' solution strategy to run two versions of each test, each copy with a different random seed.
The times reported here are the shortest time of two runs, as Z3 process terminates when either thread terminates.



\subsubsection*{Fischer Mutual Exclusion Protocol}
The well known Fischer benchmark \cite{abadi1994old} models a protocol for ensuring exclusive access to a shared common resource that can be requested by multiple processes.
The processes are identical in their behavior, aside from a numerical id, and are modeled through single TA.
The protocol uses global variables in guards and assignment statements of the TA to control access.
Each TA in the network has a `critical state', and the protocol guarantees that only one TA can be in its critical state at a time.


To measure the scalability of our new encoding, 
we performed multiple test runs while modifying the bound $k$ and the number of processes that are attempting to execute their critical region.
Several MITL properties, 
which are the same as in \cite{tack20}, were verified.
%
%
Liveness property 1 (\emph{live1}) requires that once process 1 enters state \(b\), in which it sets shared variable \(id\), it always transitions to the `waiting' state \(c\).
Property 2 (\emph{live2}) is similar, but it contains the additional constraint that process 1 must complete the transition to state \(c\) in at most 3 seconds.
Property \emph{live3} has a similar time bound, but requires that process one move to the critical section \(cs\) rather than \(c\) within the time bound, which we expect to not be universally true (a process can return to state \(b\) after moving to state \(c\) if another process has reset the variable \(id\)).
Properties \emph{live4} and \emph{live5} are copies of properties \emph{live2} and \emph{live3}, respectively, with the sole difference of inclusion vs. exclusion at the boundaries of the interval.
Property \emph{{safe}} seeks to prove the ``safety'' of the protocol, namely that two distinct processes are never in the critical section at the same time.

\begin{figure}
    \centering
    \includegraphics[width=\columnwidth]{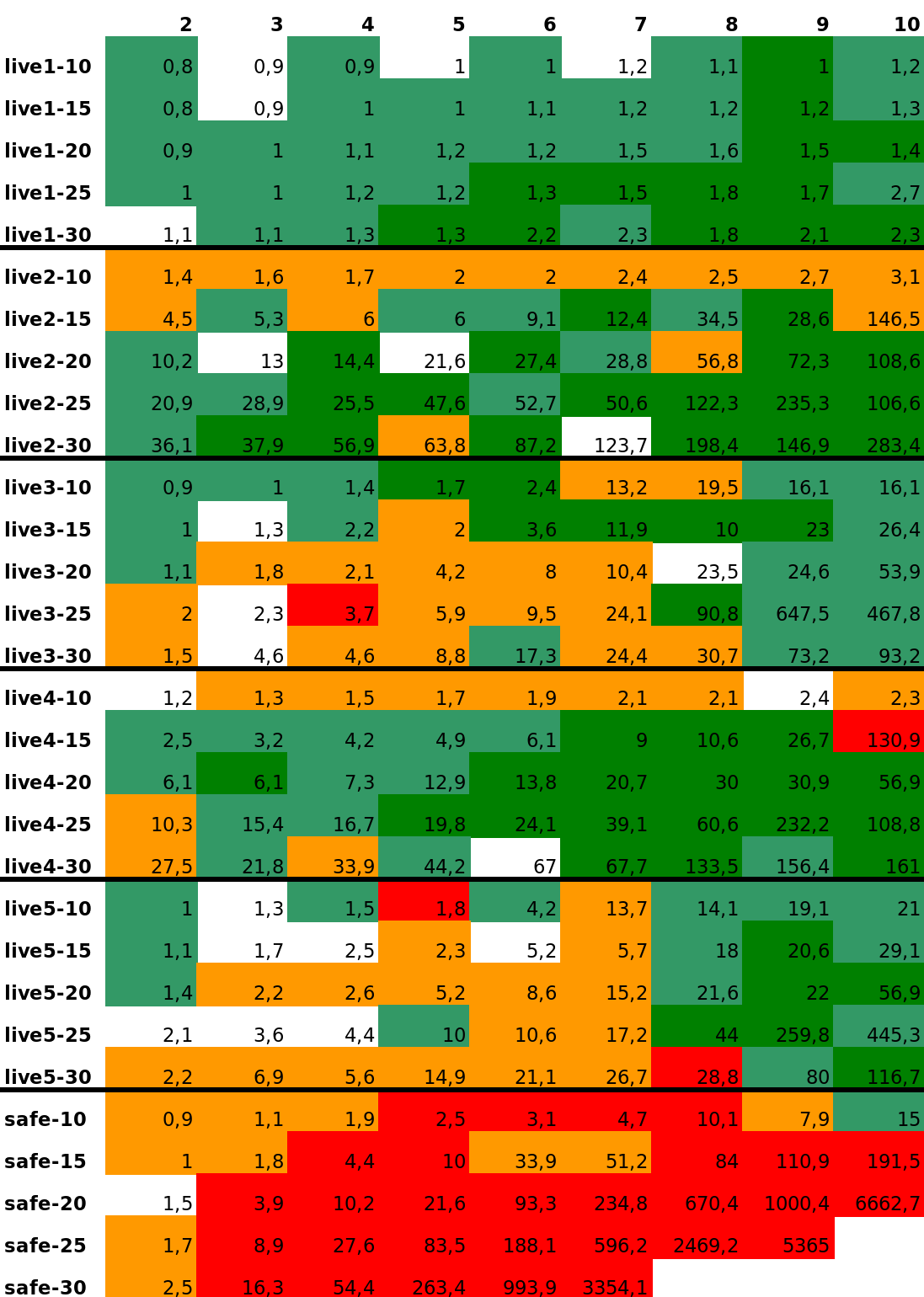}
    \caption{Results of the comparison between \newTACK{} and \oldTACK{} on the Fischer protocol. {Each column corresponds to a different number of processes involved in the Fischer protocol that share a common resource. The numbers appearing as suffixes in the property names (e.g., \emph{20} in \emph{live1-20}) indicate the length of the (lasso-shaped) traces considered by the solvers.}}
    \label{fig:fischerTACKonly}
\end{figure}

\begin{figure}
    \centering
    \includegraphics[width=\columnwidth]{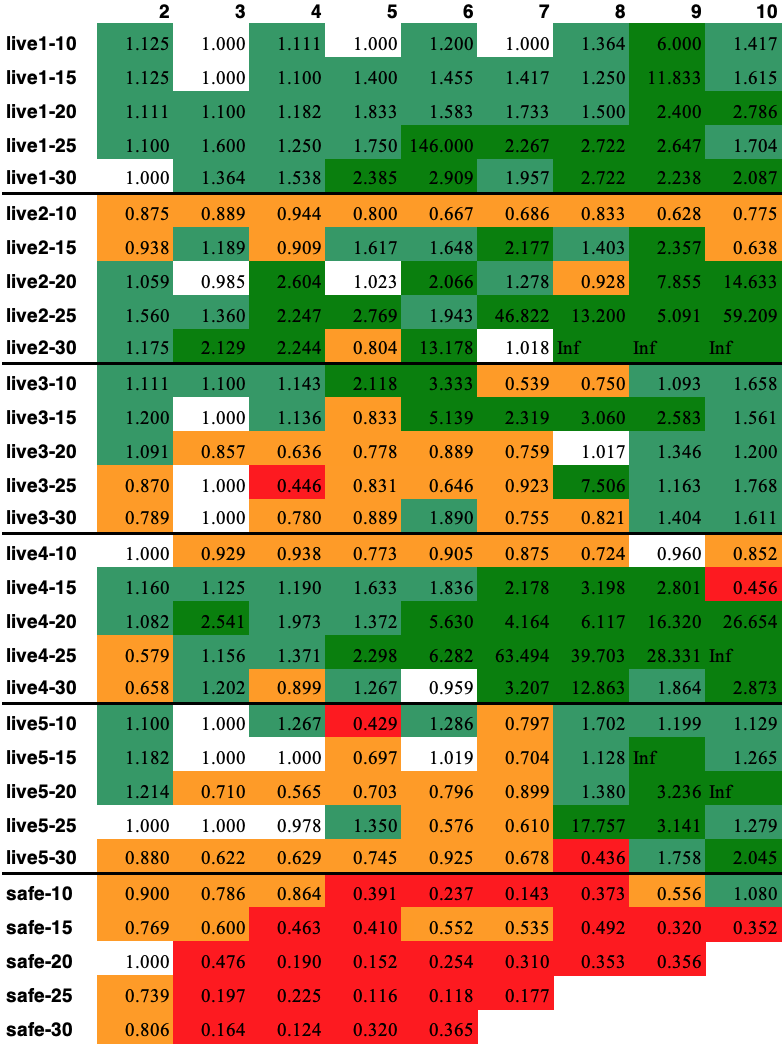}
    \caption{Speedup/slow down between \newTACK{} and \oldTACK{} on the Fischer protocol.}
    \label{fig:fischerTACKspeedup}
\end{figure}


Figure \ref{fig:fischerTACKonly} shows the results of the comparison between \newTACK{} and \oldTACK{}.
The table shows the time (sec.) that the fastest tool takes to solve the instance.
The color indicates how this time compares with 
that of \newTACK{}.
If \newTACK{} is the fastest tool of the two, the cell is colored green, and the shade of green indicates how much faster \newTACK{} is compared to \oldTACK{} (dark green means $> \mathsf{2x}$ speedup, light green means between $\mathsf{1.05x}$ and $\mathsf{2x}$ speedup).
Otherwise, the cell is colored orange when \newTACK{} is between $\mathsf{1.05x}$ and $\mathsf{2x}$ slower than \oldTACK{}, and red when it is more than $\mathsf{2x}$ slower (if the difference between \newTACK{} and \oldTACK{} is less than $5\%$ either way, the cell is left white).
Figure \ref{fig:fischerTACKspeedup} 
shows the speed up/slow down factor for each experiment with the Fischer protocol.
Empty cells indicate a timeout for both tools, set at 2 hours.
%
For property \emph{{safe}}, \oldTACK{} is the fastest tool.
Indeed, \newTACK{} is consistently faster than \oldTACK{} for greater values of the bound $k$ and higher numbers of processes, except for property \emph{{safe}}.
This property is peculiar in that the MITL formula grows in size with the number of TA in the network.
It is possible that at larger sizes, the MITL encoding becomes a bottleneck that limits the utility of further TA optimizations.

\subsubsection*{Gearbox}
\begin{figure}
    \centering
    \includegraphics[width=\columnwidth]{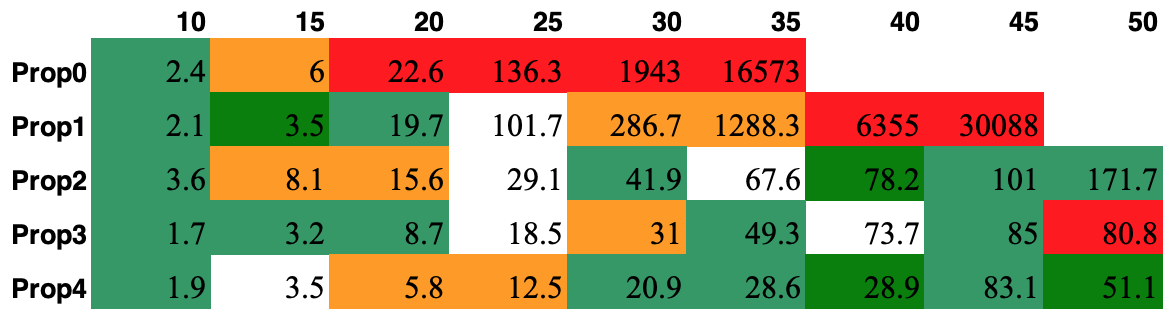}
    \caption{Results of the comparison between \newTACK{} and \oldTACK{} on the Gearbox model.}
    \label{fig:gearbox}
\end{figure}

\begin{figure}
    \centering
    \includegraphics[width=\columnwidth]{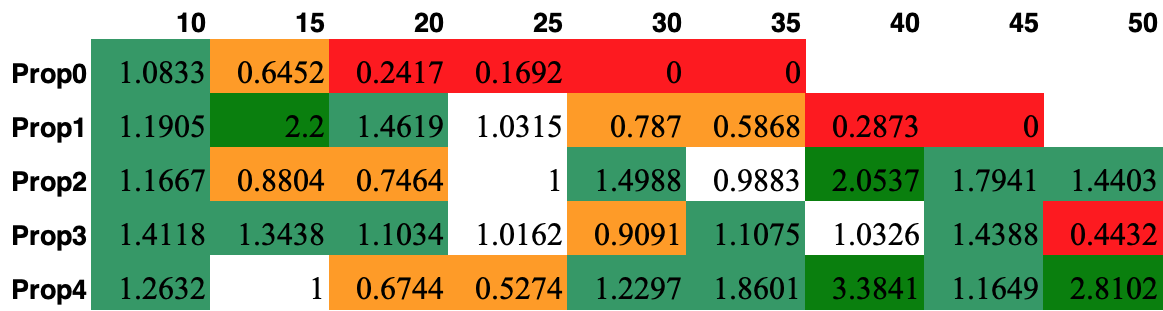}
    \caption{Speedup/slow down between \newTACK{} and \oldTACK{} on the Gearbox model.}
    \label{fig:gearboxspeedup}
\end{figure}

The Gearbox
TA models an automatic gearshift which utilizes a gearbox
controller \cite{lindahl01}. Upon receiving a gear change (reverse, neutral, as well as gears 1-5
are modelled), the controller coordinates changes to the state of the engine,
gearbox, and clutch to perform the desired gear transition. 
Property 0 asserts that in the absence of any errors, the elapsed time
required to change gears after an input is no greater than 1500 ms. Property 1
similarly asserts that for certain specific gear transitions, the absence of
errors implies a transition time of at most 1000 ms. 
Property 2 concerns error
propagation from the clutch and gearbox to the gear controller. Depending on the
specific error, the gear controller is required to respond accordingly within 200
or 350 milliseconds. 
{Property 3 also concerns error states in the controller. It asserts that each error state in the controller is active only when the related error has occurred in either the clutch or the gearbox. Thus, the controller never reports a false error.}
{Property 4 asserts that whenever the gearbox is not in neutral and no gear shift is occurring, the engine module is supplying torque to the rest of the drive system.}
These properties were evaluated over the gearbox model using
time bounds between 10 and 50 steps. 
Figure~\ref{fig:gearbox} reports, for each instance, the time (sec.) taken by the fastest tool of the two to solve the model.
As for Figure ~\ref{fig:fischerTACKonly}, cells colored green (resp., red/orange) are those for which \newTACK{} (resp., \oldTACK{}) was fastest (see Figure \ref{fig:gearboxspeedup} for the speed up/slow down factors).

\subsubsection*{Token Ring}
\begin{figure}
    \centering
    \includegraphics[width=\columnwidth]{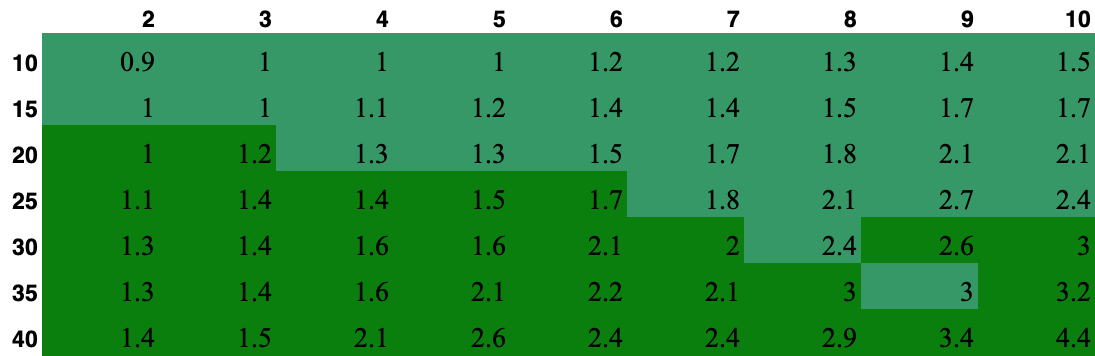}
    \caption{Results of the comparison between \newTACK{} and \oldTACK{} on the Token Ring model}
    \label{fig:tokenring}
\end{figure}

\begin{figure}
    \centering
    \includegraphics[width=\columnwidth]{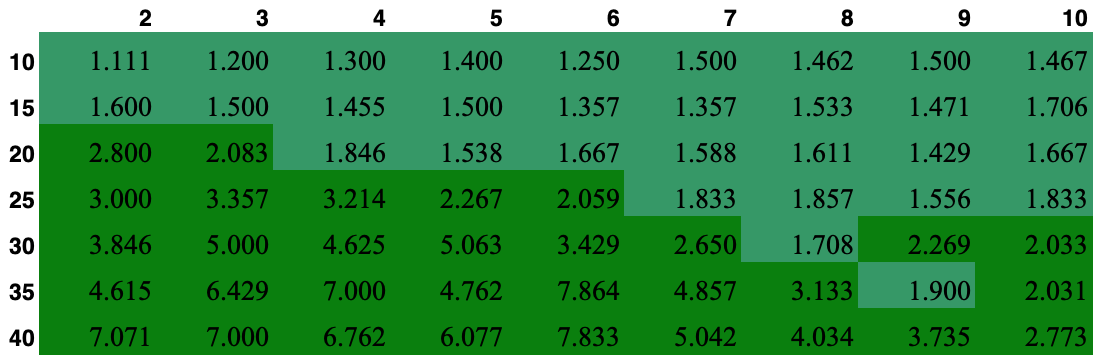}
    \caption{Speedup/slow down between \newTACK{} and \oldTACK{} on the Token Ring protocol.}
    \label{fig:tokenringspeedup}
\end{figure}

The Token Ring protocol\cite{jain1994fddi} models a ring of agents that pass a token between themselves, along with a process that models the ring itself.
The token moves in either direction along the ring (the ring process controls the token).
The agents may choose to return the token in either a synchronous or asynchronous manner.
In both cases, channel-based synchronization among TA coordinates ownership of the token.
The property checked asserts that agents 1 and 2 never simultaneously synchronize with the token.
Figure~\ref{fig:tokenring} contains the results of the Token Ring tests (time in sec.), 
while Figure \ref{fig:tokenringspeedup} shows the speed up factors.

\subsubsection*{Philips Audio Protocol}
The Philips Audio Protocol models the transmission of data over a single shared
bus between two entities. An interesting property is that the message can be decoded
by a receiver that can only detect rising edges, that is a transition from a low to
high signal over the bus. This algorithm was translated into a TA representation for the Uppal tool by Larsen et al. \cite{larsen95}. Property 1 expresses the correctness of the protocol, that with a properly functioning sender and receiver the signal will be interpreted correctly. This is represented by asserting that the receiving agent never enters an error state. Property 2 expresses that the sender will never send two rising edges within 400 units of time, regardless of the message being sent (required to ensure a 5\% timing error tolerance). Property 3 expresses that when the sender has completed the message, the receiver enters the stop state within 900 time units.
For this model, the use of arithmetic operations on variables made creating the model for \oldTACK{} (which supports such operations through a workaround) difficult, so Table~\ref{table:philips-results} shows only the results obtained with \newTACK{} on the verification of three properties.

\begin{table}
\footnotesize
\setstretch{0.9}
\center
\begin{tabular}{
r  r  r  r
r  r  r  r 
}
\toprule
  $k$  &  &    \multicolumn{1}{c}{\textbf{15}} & \multicolumn{1}{c}{\textbf{25}} & \multicolumn{1}{c}{\textbf{35}} & \multicolumn{1}{c}{\textbf{40}} & \multicolumn{1}{c}{\textbf{45}} & \multicolumn{1}{c}{\textbf{50}}  \\
  \toprule
  \multirow{3}{*}{\rotatebox[origin=c]{90}{\textbf{prop.}}}
     & \textbf{1} & 5.1 & 155.3 & 3071.9 & 9420.8 & 27947.4 & - \\
     & \textbf{2} & 3.8 & 9.0   & 29.8   & 29.8 & 42.5 & 65.0 \\
     & \textbf{3} & 7.8 & 285.8 & 6440.4 & 21426.9 & - & - \\
   \bottomrule
  \cline{1-8}\\
\end{tabular}
\caption{Time (sec.) to check the properties of the Philips protocol with \newTACK{} ($-$ means no result after 12 hours).}
\label{table:philips-results}
\end{table}

\subsubsection*{Carrier Sense Multiple Access / Collision Detection}

The CSMA/CD protocol \cite{CSMACD} is a well known protocol for allowing multiple agents to share a communication channel, and was popularized by its inclusion in the Ethernet standard.
The protocol includes one process to manage a shared communication bus, as well as a number of processes that wish to obtain exclusive access to the bus in order to send a message.
When two processes attempt to send at the same time, the bus process detects the collision and uses the broadcast synchronization primitive to force the processes to wait a randomized amount of time before attempting to communicate again.
%
The property checked 
asserts that after process 1 has been sending for 52 units of time, process 2 cannot begin sending until process 1 has finished. 
Table~\ref{table:csma-results} shows the results of the execution of the verification runs on the CSMA/CD model using \newTACK{}.
In this case, a comparison with the \oldTACK{} encoding has not been carried out, because we modified the CSMA/CD TA model to make it more accurate with respect to the real-world behavior of the protocol.
This entailed using variable comparisons that are not fully supported in \oldTACK{}, so Table~\ref{table:csma-results} only reports executions times obtained through \newTACK{}.

\begin{table}
\footnotesize
\setstretch{0.9}
\center
\begin{tabular}{
r  r  r  r
r  r  r
}
\toprule
$n$  &   &    \multicolumn{1}{c}{\textbf{3}} & \multicolumn{1}{c}{\textbf{5}} & \multicolumn{1}{c}{\textbf{7}} & \multicolumn{1}{c}{\textbf{9}} & \multicolumn{1}{c}{\textbf{10}}  \\
\toprule
  \multirow{5}{*}{\rotatebox[origin=c]{90}{$k$}}
     & \textbf{10} & 3.5    & 5.1    & 5.8    & 7.8 & 9.5 \\
     & \textbf{15} & 31.2   & 52.7   & 121.7  & 275.4 & 324.0 \\
     & \textbf{20} & 109.5  & 721.5  & 2206.1 & 5097.3 & 5068.9 \\
     & \textbf{25} & 813.7  & 4772.4 & $-$    & $-$ & $-$ \\
     & \textbf{30} & 2882.0 & $-$    & $-$    & $-$ & $-$ \\
  \bottomrule
  \cline{1-7}\\
  \end{tabular}
  \caption{Time (sec.) to check property for the CSMA/CD protocol with \newTACK{} ($-$ means no result after 2 hours).}
  \label{table:csma-results}
  \end{table}



\begin{figure}
    \centering
    \includegraphics[width=\columnwidth]{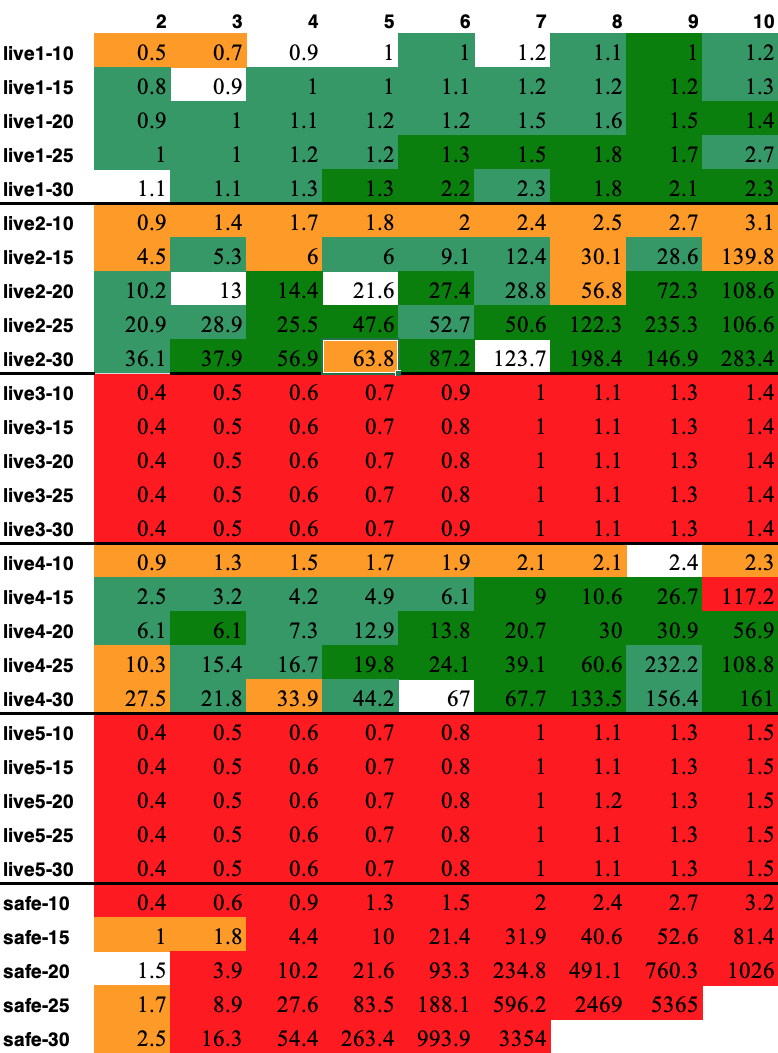}
    \caption{Results of the comparison between \newTACK{}, \oldTACK{} and \roland{} on the Fischer protocol.}
    \label{fig:fischer}
\end{figure}

\medskip
For the Fischer benchmark, in addition to the two-way comparison between \newTACK{} and \oldTACK{} discussed above and summarized by figures \ref{fig:fischerTACKonly} and \ref{fig:fischerTACKspeedup}, we also carried out a three-way comparison between the two TACK encodings and \roland{} (similarly to \cite{tack20}).
Figure \ref{fig:fischer} shows the results of the three-way comparison.
More precisely, it shows the time taken by the fastest of the three tools, and the color of each cell represents how \newTACK{} compares against the best of the other two tools: if the cell is colored green, \newTACK{} was the fastest tool, otherwise the cell is colored orange/red, with the same meaning of the coloring as for figures \ref{fig:fischerTACKonly} and \ref{fig:fischerTACKspeedup}.
Notice that the \emph{live3} and \emph{live5} properties \emph{do not hold}---i.e., a counterexample exists, as the model is \emph{satisfiable}.
In these cases, the incremental approach of \roland{}, which explores the bounds $k$ starting from $1$ until it determines that the model is satisfiable, is very efficient, since it stops the search as soon as possible.
Indeed, the portions of Figure \ref{fig:fischer} corresponding to properties \emph{live3} and \emph{live5} show that the best tool (\roland{}) uses a constant time to solve the problem, even as the bound $k$ increases (the model is satisfiable with a bound less than 10).
For property {\emph{safe}}, for higher bounds $k$, both \oldTACK{} and \newTACK{} are faster than \roland{}.

Notice that, in the cases of the Gearbox and Token Ring benchmarks, we only compared \newTACK{} against \oldTACK{}, since models for the \roland{} tool were not available (and building new ones was not possible, as explained in \cite{tack20}).

\section{Discussion and future works}
\label{sec:Concl}

Empirical testing has revealed that the novel \newTACK{} encoding can provide significant speedups across several benchmarks when compared to \oldTACK{}.
In particular, \newTACK{} is consistently better than \oldTACK{} in the Token Ring case, and mostly better in the Gearbox and Fischer cases,
especially for increasing values of the bound and of the number of processes.
These results seem to indicate that the \newTACK{} encoding is better suited to exploring models with larger bounds, as the time needed to solve larger and larger bounds grows more slowly compared to \oldTACK{}.

In addition, \newTACK{} was able to solve models (the new, more realistic CSMA/CD and the Philips protocol) that were more difficult to tackle in \oldTACK{} due to limitations in the way the old encoding deals with integer operations and with synchronizations among TA.

Future work will focus on two main objectives. Firstly, we will seek to achieve a better integration of the two translations, which will take the form of a BitVector encoding specific to MITL formulas that does not rely on the CLTLoc translation.
Secondly, given the considerable impact of incremental approaches in verification 
we will attempt to understand how to make the BitVector-based encoding incremental.

\bibliographystyle{IEEEtran}
\bibliography{bibliography}

\end{document}